\documentclass[preprint2,trackchanges]{aastex701}
\usepackage{amsmath}
\usepackage{booktabs}
\usepackage{multirow}

\begin{document}

\title{Tentative Excess Dispersion Measures toward DESI Galaxy Groups from Localized Fast Radio Bursts}

\author[orcid=0009-0007-0367-0631, gname=Jiale, sname=Wang]{Jiale Wang}
\affiliation{Department of Astronomy, Xiamen University, Xiamen 361005, People's Republic of China}
\email{wangjiale@stu.xmu.edu.cn}

\author[orcid=0000-0002-2853-3808, gname=Taotao, sname=Fang]{Taotao Fang}
\affiliation{Department of Astronomy, Xiamen University, Xiamen 361005, People's Republic of China}
\email[show]{fangt@xmu.edu.cn}
\correspondingauthor{Taotao Fang}

\begin{abstract}

Fast radio bursts (FRBs) provide sensitive probes of ionized baryons through their dispersion measures (DMs). In this work, we test whether localized FRB sight lines intersecting foreground galaxy groups exhibit a measurable excess in cosmic DM relative to clean nonintersecting sight lines. We crossmatch 117 localized FRBs with measured DMs and host redshifts against a foreground group catalog from the DESI Legacy Imaging Surveys. Within the catalog footprint and at $z_{\rm host}<0.5$, we identify 23 sight lines intersecting foreground groups and 18 clean nonintersecting sight lines for the main comparison. A joint fit to this sample gives a positive offset of $143.3^{+53.2}_{-53.0}~\mathrm{pc\,cm^{-3}}$ in the inferred cosmic DM, $\mathrm{DM}_{\rm cosmic}$, for the group-intersecting sight lines relative to the nonintersecting control sample. A template fit for the foreground halos gives $A_{\rm halo}=0.34^{+0.51}_{-0.50}$, with a positive posterior median but a weakly constrained detailed halo contribution. A mock forecast within the DESI footprint indicates that larger samples of localized FRBs would improve the foreground halo template constraint. 
\end{abstract}



\section{Introduction} 
Comparing observations of the low-redshift Universe with predictions from the standard $\Lambda$CDM cosmological model \citep{Planck2021AA...652C...4P} reveals that nearly 30\% of baryons remain unaccounted for---a discrepancy known as the ``missing baryons'' problem \citep{Shull2012}.

The prevailing view is that these missing baryons reside in hot, diffuse gas distributed throughout different environments, including the warm–hot intergalactic medium (WHIM) \citep{Dave2001ApJ...552..473D, Nicastro2018Natur.558..406N}, the circumgalactic medium (CGM) surrounding individual galaxies \citep[e.g.,][]{Nicastro2023ApJ...955L..21N}, the intragroup medium (IGrM) in galaxy groups \citep{Sullivan2017MNRAS.472.1482O,Li2024ApJ...977L..40L}, and the intracluster medium (ICM) in galaxy clusters \citep{Sun2009ApJ...693.1142S,Ge2020MNRAS.497.4704G,Ge2023MNRAS.525.1365G}.
    
However, directly quantifying the baryon content within these diffuse components remains a formidable observational challenge. Owing to their low gas densities, diffuse baryons associated with galaxies and the intergalactic medium (IGM) produce only faint X-ray and UV emission \citep{Cen2006ApJ...650..560C}, while absorption-line studies are limited to sparse lines of sight toward bright background sources \citep{Savage2009ApJ...702.1472S, Fang2015ApJS..217...21F, Luo2018ApJS..235...28L}. On larger, group- and cluster-scale halos, the thermal Sunyaev--Zel'dovich (tSZ) effect offers a complementary probe of hot gas, but its signal is typically too weak to be detected in individual systems, necessitating statistical stacking analyses \citep{Dai2007ApJ...658..917D, Anderson2015MNRAS.449.3806A, Li2024ApJ...977L..40L}.

Fast radio bursts (FRBs) are intense, millisecond-duration radio transients of predominantly extragalactic origin \citep{Lorimer2007Sci...318..777L, petroff2019fast}. As their radio pulses propagate through ionized plasma, they experience frequency-dependent arrival-time delays analogous to those observed in pulsars. This dispersion is quantified by the dispersion measure (DM), which corresponds to the integrated free-electron column density along the line of sight (LOS). Because free electrons trace the ionized baryonic component of the Universe, the DM provides a sensitive probe of ionized baryons along the line of sight, including contributions from the IGM and the gaseous halos of galaxies and groups \citep{McQuinn2014ApJ...780L..33M, Macquart2020Natur.581..391M,WANG2026SSPMA..5639601B}.

Recent observational efforts have begun to exploit FRB sight lines to probe ionized gas associated with group- and cluster-scale halos. Most recently, the CHIME/FRB Outriggers collaboration analyzed several FRBs located within or behind galaxy groups and clusters, using integrated halo density models to predict the ICM and IGrM contribution to the observed DM \citep{Lanman2025arXiv250907097L}. By comparing these model predictions with the measured DMs, they constrained the baryon fractions of individual halos and found consistency with X-ray- and tSZ-inferred relations for massive systems. This study illustrates the potential of FRBs for studying ionized gas in group and cluster environments, although the inferred constraints remain sensitive to the assumed gas-density profiles and are currently limited by small sample sizes.

In this paper, we statistically investigate whether localized FRBs exhibit excess extragalactic dispersion measures when their sight lines intersect foreground galaxy groups identified in the DESI Legacy Imaging Surveys \citep{Yang2021ApJ...909..143Y,Dey2019AJ....157..168D}. In Section~\ref{sec:Method}, we describe the FRB sample, foreground-halo identification, and the inference of cosmic dispersion measures. In Section~\ref{sec:Results}, we compare the group-intersecting and nonintersecting sight lines and model the foreground halo contribution. In Section~\ref{sec:Discussion}, we discuss the interpretation, sensitivity tests, null tests, and future-sample forecast. We summarize our conclusions in Section~\ref{sec:Conclusions}. Throughout this paper, we adopt the \citet{Planck2020A&A...641A...6P} cosmology.

\section{METHODOLOGY} \label{sec:Method}
\subsection{Localized FRB Sample}
We construct a parent sample of 117 localized FRBs with measured DMs, spectroscopic host redshifts, and localization uncertainties below $20''$. The adopted sky positions, localization uncertainties, observed DMs, Galactic DM estimates, host redshifts, and localization references are listed in Table~\ref{tab:frb-sample}.

\begin{figure*}
    \centering
    \includegraphics[width=\textwidth]{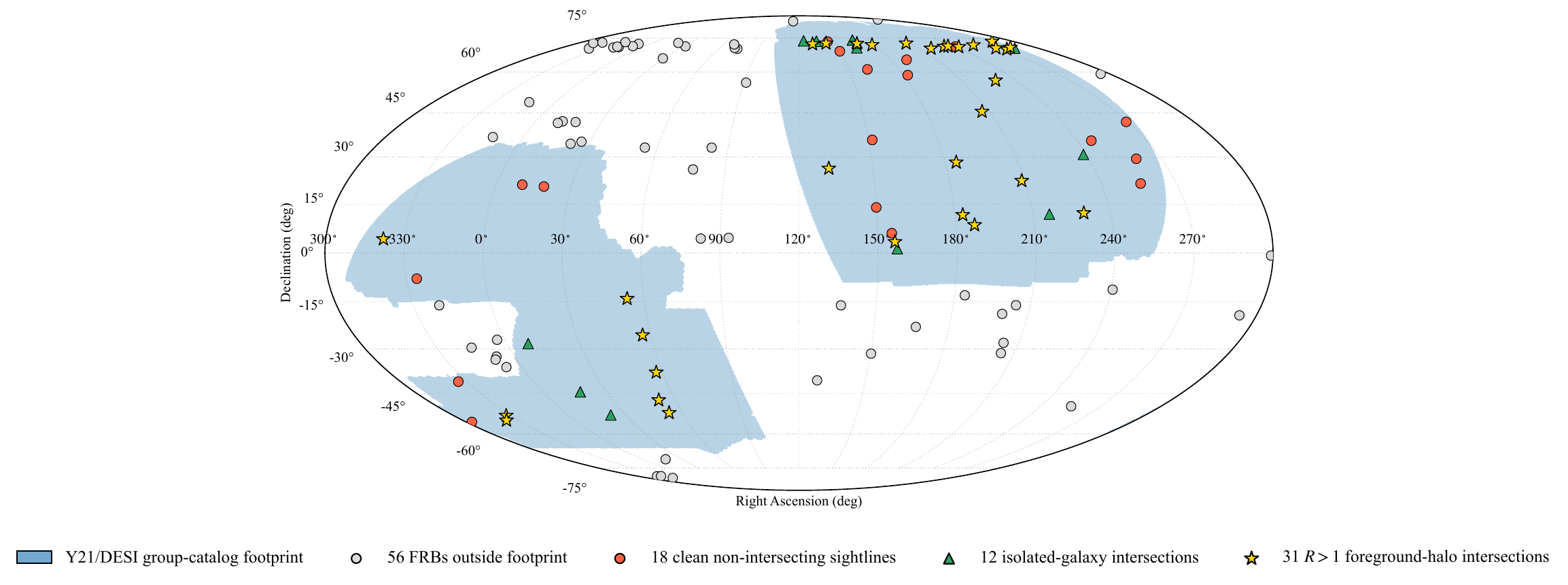}
    \caption{
Sky distribution of the 117 localized FRBs in the parent sample, shown in equatorial coordinates (RA, Dec). The blue-shaded regions indicate the sky coverage of the foreground group catalog used in this work. Red circles mark FRBs inside the group-catalog footprint with no identified foreground-halo intersection, golden stars mark sight lines intersecting foreground halos with more than one identified member ($R>1$), and green triangles denote sight lines intersecting only isolated galaxy-scale halos ($R=1$). Gray circles represent FRBs outside the group-catalog footprint.
}
    \label{fig:sky_distribution}
\end{figure*}

\subsection{Foreground Halo Identification Using the DESI Group Catalog}\label{sec:Identification}
For the identification of foreground halos intersected by the FRB sight lines, we construct our foreground-system sample from the optically selected catalog of \citet[][hereafter Y21]{Yang2021ApJ...909..143Y}, which is built using data from the DESI Legacy Imaging Surveys \citep{Dey2019AJ....157..168D}. The catalog is generated using an updated halo-based group finder \citep{Yang2005MNRAS.356.1293Y, Yang2007ApJ...671..153Y} applied to the Data Release 9 (DR9) galaxy sample.

We crossmatch the localized FRBs with this group catalog to identify intervening halos along each sight line. A system is classified as an intersecting foreground halo when the projected impact parameter $b$, defined as the transverse separation between the FRB sight line and the group center, satisfies $b<r_{180}$. Following the radius definition adopted in Y21, we use
\begin{equation}
\begin{aligned}
r_{180} &= 0.78\,h^{-1}\,{\rm Mpc} \\
&\quad \times
\left( \frac{M_h}{\Omega_M 10^{14}\,h^{-1}\,M_\odot} \right)^{1/3}
(1+z_{\rm group})^{-1},
\end{aligned}
\end{equation}
which corresponds to an enclosed overdensity of 180 relative to the cosmic mean matter density.

To separate foreground groups from the FRB host environment, we adopt a redshift selection criterion that explicitly accounts for redshift uncertainties. The Y21 group catalog contains both spectroscopic- and photometric-redshift systems. While redshift information and quality flags are provided at the galaxy level, the group catalog itself lists only a single group redshift and does not distinguish between spectroscopic- and photometric-redshift systems. We therefore infer the redshift type of each group from its member galaxies: a group is classified as spectroscopic only if all its member galaxies have spectroscopic redshifts; otherwise, it is treated as photometric.

For photometric-redshift groups, the typical uncertainty can be approximated as $\sigma_{z,\mathrm{photo}} \simeq 0.01 + 0.015z$, whereas spectroscopic-redshift groups have much smaller uncertainties, with a typical value of $\sigma_{z,\mathrm{spec}} \sim 10^{-4}$. This wide range of redshift uncertainties motivates a redshift-selection scheme that adapts to the differing precision of spectroscopic and photometric measurements.

Our approach is informed by the strategy of \citet[][hereafter M25]{Mo2025ApJ...995..183M}, who adopted a fixed redshift offset of $\Delta z = 0.05$ matched to the typical uncertainty of the DESI photometric-redshift galaxy catalog used in their analysis, for which $\sigma_z \sim 0.02$ \citep{Zou2019ApJS..242....8Z}. However, applying the same fixed redshift offset to the Y21 catalog would be overly conservative for most systems and could exclude genuine foreground halos.

We therefore adopt a $\sigma$-scaled redshift separation criterion,
\begin{equation}
z_{\rm FRB} \ge z_{\rm group} + 2.5\,\sigma_z,
\end{equation}
where $\sigma_z$ is set to $\sigma_{z,\mathrm{spec}}$ or $\sigma_{z,\mathrm{photo}}$ depending on the inferred redshift type of the group. We adopt $2.5\sigma_z$ as a conservative threshold for identifying foreground systems. We have verified that lowering this threshold to values between $1.0\sigma_z$ and $2.5\sigma_z$ changes only a small number of foreground-intersection classifications.

For systems with richness $R = 1$ in the Y21 group catalog, only a single member galaxy is identified within the associated dark matter halo. For the purpose of our analysis, these systems are treated as galaxy-dominated halos, for which the group-scale boundary $r_{180}$ likely extends well beyond the spatial extent of the CGM associated with an individual galaxy. We therefore adopt the virial radius, $r_{\rm vir}$, as a more appropriate characteristic scale for these systems. The corresponding virial mass, $M_{\rm vir}$, and radius, $r_{\rm vir}$, are obtained by converting the Y21 halo mass defined at $r_{180}$, assuming a Navarro–Frenk–White (NFW) profile \citep{Navarro1997ApJ...490..493N} with halo concentrations given by the mass–concentration relation of \citet{Diemer2019ApJ...871..168D}. Accordingly, an FRB sight line is classified as intersecting such a galaxy-dominated halo when its projected impact parameter satisfies $b < r_{\rm vir}$.

For the main comparison, we define foreground groups as Y21 systems with richness $R>1$. sight lines intersecting such systems are classified as group-intersecting sight lines. This definition also includes a small number of massive systems in the cluster-scale regime, which we identify as systems with $M_{500}\gtrsim10^{14}M_\odot$, a commonly used approximate boundary between galaxy groups and clusters in hot-gas studies \citep[e.g.,][]{Sun2009ApJ...693.1142S}. We retain these systems as the high-mass tail of the same foreground-system sample.

Figure~\ref{fig:sky_distribution} shows the sky distribution and foreground classifications of the 117 FRBs. The identified foreground systems are listed in Table~\ref{tab:frb_groups}.

For the main comparison between group-intersecting and nonintersecting sight lines, we further restrict the footprint sample to FRBs with $z_{\rm host}<0.5$, where the Y21 group catalog provides a more uniform basis for foreground group identification \citep{Yang2021ApJ...909..143Y}. After this redshift cut, the footprint sample contains 23 group-intersecting sight lines (Pass) and 18 clean nonintersecting sight lines (NoPass) for the main comparison. sight lines intersecting only isolated galaxy halos are shown separately in Figure~\ref{fig:sky_distribution}, but are excluded from the main comparison because such single-galaxy intersections are more sensitive to localization uncertainties than group-scale intersections.

Over the redshift range of the main comparison, the adopted $20''$ localization threshold corresponds to a projected physical scale of $\lesssim130~{\rm kpc}$, which is small compared with the characteristic radii of the group- and cluster-scale foreground systems used for the Pass classification. {We further quantify the effects of FRB localization and foreground-system redshift uncertainties through the sensitivity tests presented in Section~\ref{sec:robustness}.}

\subsection{Inferred Cosmic Dispersion Measures}\label{sec:DMcosmic}
We define the inferred cosmic DM, $\mathrm{DM}_{\rm cosmic}$, by subtracting the Milky Way disk, Milky Way halo, and host-galaxy contributions from the observed DM of an FRB:
\begin{equation}\label{eq:dmcosmic}
\begin{aligned}
\mathrm{DM}_{\rm cosmic}(z)
&= \mathrm{DM}_{\rm obs}
 - \mathrm{DM}_{\rm MW,disk} \\
&\quad
 - \mathrm{DM}_{\rm MW,halo}
 - \frac{\mathrm{DM}_{\rm host}}{1+z}.
\end{aligned}
\end{equation}
Here, $\mathrm{DM}_{\rm MW,disk}$, $\mathrm{DM}_{\rm MW,halo}$, and $\mathrm{DM}_{\rm host}$ denote the contributions from the Milky Way disk, the Milky Way halo, and the FRB host galaxy, respectively. The factor $(1+z)^{-1}$ converts the host-frame DM contribution to the observer-frame contribution.

For the Milky Way disk, Galactic electron-density models such as NE2001 \citep{Cordes2002astro.ph..7156C} and YMW16 \citep{Yao2017ApJ...835...29Y} are commonly used to estimate $\mathrm{DM}_{\rm MW,disk}$. In this work, we adopt the recently updated NE2025 model \citep{Ocker2026ApJ..1002....3O}, which extends the NE2001 framework using an expanded set of pulsar dispersion and scattering measurements, to set the mean value of $\mathrm{DM}_{\rm MW,disk}$. For the Milky Way halo, previous studies often adopted a fixed contribution \citep[e.g.,][]{Macquart2020Natur.581..391M,Batten2021MNRAS.505.5356B,Connor2025NatAs...9.1226C}. We instead use the YT2020 model of \citet{Yamasaki2020ApJ...888..105Y}, which predicts a direction-dependent halo DM based on X-ray constraints and is therefore better suited for estimating the halo contribution toward individual FRBs. This choice is also motivated by recent studies indicating that the Milky Way halo contribution can be anisotropic \citep{Wang2025Univ...11...41W,Huang2025MNRAS.538.2785H}. Following \citet{Sang2025MNRAS.544.2870S}, both Galactic components are modeled as Gaussian distributions with a conservative fractional uncertainty of $\sigma=0.5\,\mu$.

For the host contribution, $\mathrm{DM}_{\rm host}$, we adopt the same lognormal form as M25 and \citet{Connor2025NatAs...9.1226C}, with the median set by stellar mass when available. For FRBs with measured host stellar masses, we set the lognormal median, $\widetilde{\mathrm{DM}}_{\rm host}=\exp(\mu_{\rm host})$, using the empirical relation of \citet{Bernales2025A&A...696A..81B},
\begin{equation}
\exp(\mu_{\rm host})
=
\left[
43\,\log_{10}\!\left(\frac{M_\star}{10^{10}\,M_\odot}\right)
+ 86
\right]~\mathrm{pc\,cm^{-3}} .
\end{equation}
The stellar masses are taken from the compilation of \citet{Sang2025MNRAS.544.2870S} (see their Table~A1). For FRBs without reliable stellar-mass estimates, we use the fiducial prescription adopted by M25 and \citet{Connor2025NatAs...9.1226C}, namely a fixed median $\exp(\mu_{\rm host})=130~\mathrm{pc\,cm^{-3}}$ and logarithmic width $\sigma_{\rm host}=0.56$. 

Combining these distributions, we infer the posterior distribution of $\mathrm{DM}_{\rm cosmic}$ for each FRB by marginalizing over the Milky Way foreground and host-galaxy contributions:
\begin{equation}
\begin{split}
p_{\rm cosmic}(\mathrm{DM}_{\rm cosmic})
&=
\mathcal{N}^{-1}
\int_{0}^{\infty}
p_G(\Delta\mathrm{DM}) \\
&\quad \times
p_{\rm host}(\mathrm{DM}_{\rm host})\,
\mathrm{dDM}_{\rm host}.
\end{split}
\end{equation}
Here $p_{\rm G}$ and $p_{\rm host}$ denote the Galactic and host DM distributions, respectively, $\Delta\mathrm{DM}\equiv\mathrm{DM}_{\rm obs}-\mathrm{DM}_{\rm cosmic}-\mathrm{DM}_{\rm host}/(1+z)$, and $\mathcal{N}$ is a normalization constant. For physical consistency, we require $\mathrm{DM}_{\rm cosmic}\ge0$.

{For each FRB $i$, we report the posterior median of $\mathrm{DM}_{{\rm cosmic},i}$ and define its symmetrized uncertainty as}
\begin{equation}
\sigma_{{\rm cosmic},i}
=
\frac{1}{2}
\left[
\mathrm{DM}_{{\rm cosmic},i}^{(84)}
-
\mathrm{DM}_{{\rm cosmic},i}^{(16)}
\right],
\label{eq:sigma-cosmic}
\end{equation}
{where $\mathrm{DM}_{{\rm cosmic},i}^{(16)}$ and $\mathrm{DM}_{{\rm cosmic},i}^{(84)}$ denote the 16th and 84th percentiles of the individual posterior, respectively. The corresponding values of $\sigma_{{\rm cosmic},i}$ are also listed in Table~\ref{tab:frb-sample}.}

\subsection{{Statistical Analysis of the Pass and NoPass Samples}}
\label{sec:statistical-analysis}

{We use these inferred $\mathrm{DM}_{\rm cosmic}$ values and their uncertainties to compare the Pass and NoPass samples. We fit all sight lines simultaneously with a common linear redshift relation and an additional mean offset for Pass sight lines:}
\begin{equation}
\mathrm{DM}_{{\rm cosmic},i}
=
\alpha z_i+\Delta_{\rm Pass}P_i+\epsilon_i ,
\label{eq:joint-offset}
\end{equation}
{where $P_i=1$ for Pass sight lines and $P_i=0$ for NoPass sight lines. We model the individual $\mathrm{DM}_{{\rm cosmic},i}$ values as independent Gaussian variates with mean $\mu_i=\alpha z_i+\Delta_{\rm Pass}P_i$ and variance $s_i^2=\sigma_{{\rm cosmic},i}^{2}+\sigma_{\rm int}^{2}$. A single $\alpha$ and $\sigma_{\rm int}$ are shared by both samples, while $\Delta_{\rm Pass}$ quantifies the mean offset of the Pass sample relative to the NoPass relation at fixed redshift. The resulting joint log-likelihood is}
\begin{equation}
\begin{aligned}
\ln\mathcal{L}
=
-\frac{1}{2}\sum_i
\bigg[
&\frac{
\left(\mathrm{DM}_{{\rm cosmic},i}-\mu_i\right)^2
}{s_i^2}\\
&+\ln\!\left(2\pi s_i^2\right)
\bigg].
\end{aligned}
\label{eq:joint-offset-likelihood}
\end{equation}

{We jointly sample $\alpha$, $\Delta_{\rm Pass}$, and $\sigma_{\rm int}$ using \texttt{emcee} \citep{Foreman-Mackey2013PASP..125..306F}. We adopt uniform priors over $0<\alpha<3000$ and $-3000<\Delta_{\rm Pass}<3000$, and a log-uniform prior over $10^{-3}<\sigma_{\rm int}<10^{4}$. We use 48 walkers and 8,000 steps with a random seed of 42, discard the first 2,000 steps, and retain every fifth postburn sample. The resulting joint posterior distributions and parameter covariances are shown in Figure~\ref{fig:joint-offset-corner}.}

{As an additional sensitivity check, we repeat the fit with a separate residual-scatter parameter for each sample. The mean model and priors on $\alpha$ and $\Delta_{\rm Pass}$ remain unchanged; details and results are presented in Appendix~\ref{app:separate-scatter}.}

\section{Results} \label{sec:Results}
\subsection{$\mathrm{DM}_{\rm cosmic}$--$z$ comparison of the Pass and NoPass samples}
\label{subsec:DMcosmic-z}

\begin{figure*}[ht]
    \centering
    \includegraphics[width=1.0\linewidth]{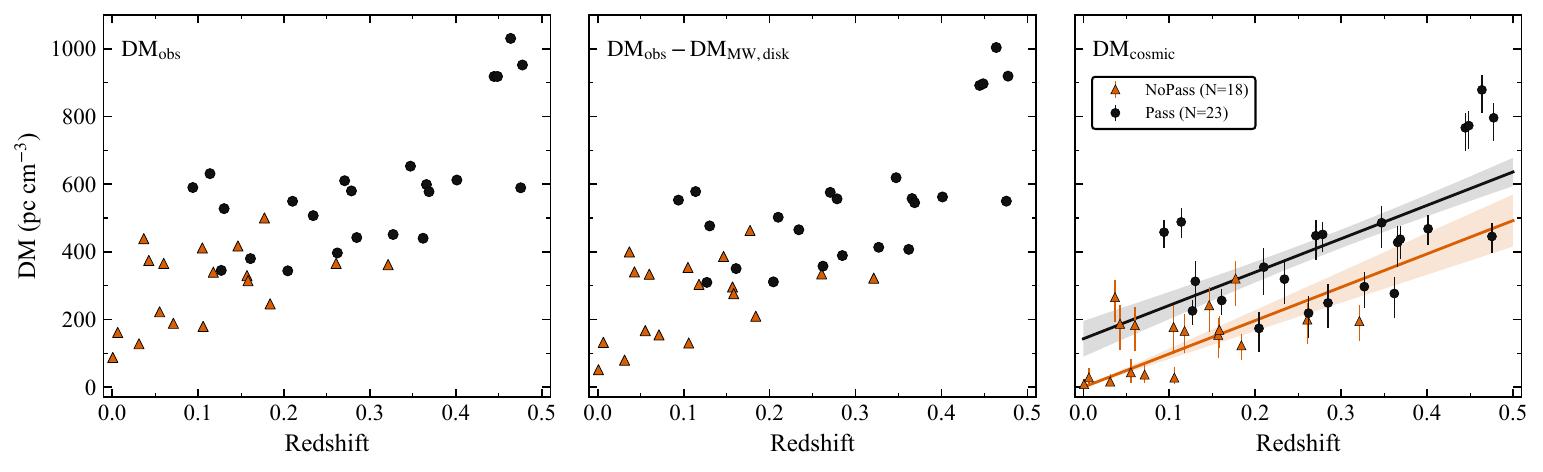}
    \caption{
   Comparison of the Pass and NoPass samples with $z_{\rm host}<0.5$ using three progressively processed DM quantities as a function of redshift. The left, middle, and right panels show $\mathrm{DM}_{\rm obs}$, $\mathrm{DM}_{\rm obs}-\mathrm{DM}_{\rm MW,disk}$ after subtracting the Milky Way disk contribution, and $\mathrm{DM}_{\rm cosmic}$ after further accounting for the Milky Way halo and host-galaxy contributions, respectively. Black circles denote group-intersecting sight lines (Pass, $N=23$), and orange triangles denote clean nonintersecting sight lines (NoPass, $N=18$). In the right panel, vertical error bars show the 16th--84th percentile ranges of the individual $\mathrm{DM}_{\rm cosmic}$ posteriors, while the solid lines and shaded bands show the posterior medians and 16th--84th percentile ranges of the fitted relations.
}
    \label{fig:DMcosmic-z}
\end{figure*}

Figure~\ref{fig:DMcosmic-z} compares the group intersecting and non intersecting samples using $\mathrm{DM}_{\rm obs}$, $\mathrm{DM}_{\rm obs}-\mathrm{DM}_{\rm MW,disk}$, and $\mathrm{DM}_{\rm cosmic}$. Pass sight lines tend to have larger DMs at a given redshift. Removing the Milky Way disk contribution leaves this pattern largely unchanged, indicating that the Galactic disk is not the main source of the difference.

{Applying the joint model defined by
Equations~\ref{eq:joint-offset} and
\ref{eq:joint-offset-likelihood}, we obtain}
\begin{equation}
\alpha =
984.5^{+154.6}_{-155.5}~\mathrm{pc\,cm^{-3}},
\end{equation}
and
\begin{equation}
\Delta_{\rm Pass}
=
143.3^{+53.2}_{-53.0}~\mathrm{pc\,cm^{-3}},
\end{equation}
with
\begin{equation}
\sigma_{\rm int}
=
117.6^{+17.5}_{-14.8}~\mathrm{pc\,cm^{-3}} .
\end{equation}

{The joint posterior gives $P(\Delta_{\rm Pass}>0)=0.9944$, equivalent to $2.54\sigma$ on a one-sided Gaussian scale, supporting a higher mean $\mathrm{DM}_{\rm cosmic}$ for group-intersecting sight lines than for clean nonintersecting sight lines.}

{We also examine the Pass--NoPass difference through comparisons of the regression coefficients, the intercept offset under a common-slope model, and the slope-removed distributions (see Appendix~\ref{app:complementary-tests}).}

\subsection{$\mathrm{DM}_{\rm excess}$ for Group-intersecting sight lines} \label{sec:3.2}

To quantify the excess for each Pass sight line, we define $\mathrm{DM}_{\rm excess}$ relative to the NoPass baseline implied by Equation~\ref{eq:joint-offset},
\begin{equation}
\mathrm{DM}_{{\rm excess},i}
=
\mathrm{DM}_{{\rm cosmic},i}
-
\alpha z_i ,
\end{equation}
where $\alpha z_i$ is the fitted NoPass relation at the same redshift.

\begin{deluxetable}{lccc}
\tabletypesize{\footnotesize}
\tablewidth{\columnwidth}
\tablecaption{sight line $\mathrm{DM}_{\rm excess}$ values for group-intersecting FRB sight lines, defined relative to the NoPass baseline implied by the joint offset fit, together with the corresponding modeled foreground-halo DM contributions.\label{tab:dmexcess_pass}}
\tablehead{
\colhead{FRB} &
\colhead{$z_{\rm FRB}$} &
\colhead{$\mathrm{DM}_{\rm excess}$} &
\colhead{$\mathrm{DM}_{\rm model}$} \\
\colhead{} &
\colhead{} &
\colhead{(pc cm$^{-3}$)} &
\colhead{(pc cm$^{-3}$)}
}
\startdata
181112A & 0.4755 & $-22.5\pm85.7$ & $ 238.4\pm32.4$ \\
191001A & 0.2340 & $ 88.7\pm72.8$ & $ 15.3\pm2.9$ \\
200430A & 0.1610 & $ 97.7\pm45.8$ & $ 44.1\pm11.1$ \\
200906A & 0.3688 & $ 74.9\pm75.2$ & $ 112.7\pm43.9$ \\
220105A & 0.2785 & $ 177.6\pm62.2$ & $ 11.1\pm2.6$ \\
220204A & 0.4012 & $ 73.2\pm76.4$ & $ 133.2\pm42.9$ \\
220914A & 0.1139 & $ 376.0\pm46.6$ & $ 46.4\pm8.3$ \\
221012A & 0.2847 & $-31.4\pm78.2$ & $ 84.6\pm10.4$ \\
221106A & 0.2044 & $-27.5\pm67.8$ & $ 3.9\pm2.0$ \\
230124A & 0.0939 & $ 365.6\pm44.2$ & $ 19.7\pm4.8$ \\
230307A & 0.2706 & $ 181.4\pm72.5$ & $ 187.9\pm26.0$ \\
230616A & 0.4485 & $ 331.5\pm89.3$ & $ 73.9\pm11.3$ \\
230626A & 0.3270 & $-25.2\pm73.1$ & $ 11.5\pm3.0$ \\
230628A & 0.1270 & $ 100.3\pm42.3$ & $ 3.9\pm1.6$ \\
230702A & 0.3658 & $ 67.7\pm83.4$ & $ 23.0\pm7.9$ \\
230808F & 0.3472 & $ 144.1\pm81.1$ & $ 10.0\pm2.5$ \\
230902A & 0.3619 & $-79.4\pm81.6$ & $ 41.9\pm8.1$ \\
230907D & 0.4638 & $ 422.2\pm91.1$ & $ 133.5\pm19.1$ \\
230923A & 0.4445 & $ 329.2\pm88.7$ & $ 76.9\pm11.8$ \\
231020B & 0.4775 & $ 326.3\pm92.9$ & $ 65.0\pm9.1$ \\
231123B & 0.2621 & $-39.3\pm74.2$ & $ 87.8\pm41.6$ \\
240114A & 0.1303 & $ 184.5\pm76.1$ & $ 109.7\pm25.2$ \\
240215A & 0.2100 & $ 147.8\pm75.8$ & $ 3.8\pm1.7$ \\
\enddata
\tablecomments{$\mathrm{DM}_{\rm excess}$ is measured relative to the NoPass baseline from the joint offset fit. $\mathrm{DM}_{\rm model}$ is the median summed foreground halo contribution from the model grid described in Section~\ref{sec:DMmodel}. Quoted uncertainties are symmetrized half-widths of the 16th--84th percentile intervals.}
\end{deluxetable}

The resulting $\mathrm{DM}_{\rm excess}$ values are listed in Table~\ref{tab:dmexcess_pass}. Their uncertainties are obtained by Monte Carlo propagation of the $\mathrm{DM}_{\rm cosmic}$ uncertainties and the posterior samples of $\alpha$.

\subsection{Modeling Foreground Halo DM Contributions}\label{sec:DMmodel}

Observational evidence indicates that galaxies, groups, and clusters are surrounded by substantial reservoirs of hot, ionized gas in the CGM, IGrM, and ICM \citep[e.g.,][]{Tiwari2021MNRAS.500.5524T,Sun2012NJPh...14d5004S,Gastaldello2021Univ....7..208G}. To examine whether the foreground intersections can account for the measured $\mathrm{DM}_{\rm excess}$, we estimate the modeled halo contribution, $\mathrm{DM}_{\rm model}$, along each Pass sight line.

Some Pass sight lines also contain intersections with $R=1$ systems, as listed in Table~\ref{tab:frb_groups}. These $R=1$ systems are not used to define the Pass sample, but their diffuse hot halo gas contribution is included in $\mathrm{DM}_{\rm model}$.

For the diffuse ionized halo gas, we adopt a spherical $\beta$ model for all foreground halos \citep[e.g.,][]{Croston2008A&A...487..431C}. The gas density profile is written as
\begin{equation}
\rho_{\rm g}(r) =
\rho_0 \left[1+\left(\frac{r}{r_{\rm c}}\right)^2\right]^{-3\beta/2},
\end{equation}
where $\rho_0$ is the normalization, $r_{\rm c}$ is the core radius, and $\beta$ determines the outer slope of the profile.

For galaxy-scale halos, we adopt $0.35 \le \beta \le 0.53$, motivated by the eROSITA stacking analysis of \citet{Zhang2024A&A...690A.267Z}. That study finds $\beta \approx 0.43^{+0.10}_{-0.06}$ for Milky Way--mass systems and $\beta \approx 0.37^{+0.04}_{-0.02}$ for M31-mass systems, with comparable values in the higher-mass bin. We then follow the subsequent modeling framework of \citet{Zhang2026A&A...706A.102Z}, which builds on these eROSITA measurements, and adopt $1 \le r_{\rm c}/{\rm kpc} \le 5$ together with a hot-gas normalization $M_{\rm hot}=f_{\rm hot}M_{\rm vir}$, where $0.03 \le f_{\rm hot} \le 0.10$.

For group-scale halos, we adopt $0.35\leq\beta\leq{0.51}$ and $0.05\leq r_{\rm c}/R_{500}\leq0.10$, while for cluster-scale halos we adopt $0.60\leq\beta\leq{0.81}$ and $0.08\leq r_{\rm c}/R_{500}\leq0.15$. These ranges are chosen to cover plausible gas-density profiles for group- and cluster-scale systems and are guided by representative X-ray constraints on hot gas in groups and clusters \citep[e.g.,][]{Ge2016MNRAS.459..366G,Sullivan2017MNRAS.472.1482O,Li2024ApJ...977L..40L,Croston2008A&A...487..431C}.

Following \citet{Akino2022PASJ...74..175A}, we adopt the empirical $M_{500}$--$f_{\rm gas,500}$ relation of \citet{Eckert2016A&A...592A..12E} to normalize the hot-gas content of group- and cluster-scale halos:
\begin{equation}
h_{70}^{-3/2} f_{\rm gas,500}
=
0.055^{+0.007}_{-0.006}
\left(
\frac{M_{500}}{10^{14}h_{70}^{-1}M_\odot}
\right)^{0.21^{+0.11}_{-0.10}} .
\end{equation}

For each group- or cluster-scale halo, the mass defined at $r_{180}$ is converted to $M_{500}$ using the halo mass--overdensity conversion of \citet{Diemer2019ApJ...871..168D}. We then set $M_{\rm gas,500}=f_{\rm gas,500}M_{500}$ and fix the density normalization $\rho_0$ by requiring that the enclosed gas mass within $R_{500}$ reproduces this value, while keeping the profile shape unchanged.

For all halos, the DM contribution is computed by integrating the corresponding electron density profile along the FRB sight line and applying the standard $(1+z)^{-1}$ cosmological correction at the halo redshift. For galaxy-scale halos the line-of-sight integration is performed out to $R_{\rm vir}$, while for group- and cluster-scale halos it is performed out to $r_{180}$. We assume fully ionized gas with primordial composition, corresponding to a mean molecular weight per free electron of $\mu_e = 1.17$.

Previous studies have often adopted modified NFW (mNFW) profiles to describe hot gas in galaxy groups and clusters \citep[e.g.,][]{Prochaska2019MNRAS.485..648P,Faber2024arXiv240514182F,Huang2025MNRAS.538.2785H,Lanman2025arXiv250907097L}. Our fiducial choice is motivated by recent kinematic Sunyaev--Zel'dovich measurements suggesting that gas in group-scale halos may be more spatially extended than the underlying dark-matter halo \citep{Hadzhiyska2025PhRvD.112h3509H}, and by the flexibility of the $\beta$ model in representing extended gas distributions consistent with existing X-ray constraints.

For each Pass sight line, we compute the DM contribution of each identified foreground halo over the adopted grids of the $\beta$-model parameters and gas normalizations. For sight lines with multiple intersections, independent draws from the corresponding halo grids are summed to construct the distribution of the total modeled foreground DM. We define the median of this distribution as $\mathrm{DM}_{{\rm model},i}$ and use the symmetrized 16th--84th percentile range as $\sigma_{{\rm DM,model},i}$. These quantities are also listed in Table~\ref{tab:dmexcess_pass}.

\subsection{Foreground-halo template fit}
We next fit the amplitude of the modeled foreground halo contribution jointly with the baseline $\mathrm{DM}_{\rm cosmic}$--$z$ relation, using
\begin{equation}
\mu_i = \alpha_{\rm h} z_i + A_{\rm halo} H_i,
\end{equation}
where
\begin{equation}
H_i =
\begin{cases}
\mathrm{DM}_{{\rm model},i}, & \mathrm{Pass},\\
0, & \mathrm{NoPass}.
\end{cases}
\end{equation}
The likelihood variance is
\begin{equation}
\sigma_{{\rm tot},i}^2 =
\sigma_{{\rm cosmic},i}^2
+
A_{\rm halo}^2 \sigma_{H,i}^2
+
\sigma_{\rm int,h}^2 ,
\end{equation}
where $\sigma_{H,i}=\sigma_{{\rm DM,model},i}$ for Pass sight lines and $\sigma_{H,i}=0$ for NoPass sight lines. {The coefficient $A_{\rm halo}$ scales the adopted foreground-halo template, with $A_{\rm halo}=1$ corresponding to the nominal template prediction and $A_{\rm halo}=0$ to no template contribution. Because the model does not include a separate Pass-sample offset, $A_{\rm halo}$ can covary with $\alpha_{\rm h}$ and is not an independent constraint on the halo-gas normalization.} The parameters $\alpha_{\rm h}$ and $\sigma_{\rm int,h}$ describe the redshift trend and residual scatter, respectively.

\begin{figure}[t]
    \centering
    \includegraphics[width=\linewidth]{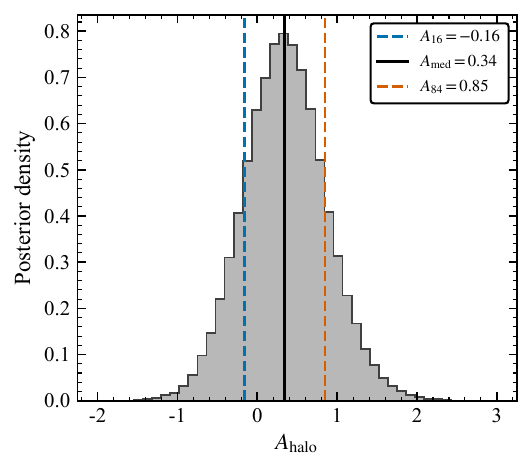}
    \caption{
Posterior distribution of the foreground-halo template amplitude $A_{\rm halo}$. The blue dashed, black solid, and orange dashed vertical lines mark the 16th percentile, median, and 84th percentile of the posterior distribution, corresponding to $A_{16}=-0.16$, $A_{\rm med}=0.34$, and $A_{84}=0.85$, respectively.
}
    \label{fig:A_posterior}
\end{figure}

The priors on $\alpha_{\rm h}$ and $\sigma_{\rm int,h}$ are the same as those used above, with a uniform prior over $-10<A_{\rm halo}<10$. The fit gives $A_{\rm halo}=0.34^{+0.51}_{-0.50}$, with $P(A_{\rm halo}>0)=0.751$. As shown in Figure~\ref{fig:A_posterior}, the posterior median is positive, but the 16th--84th percentile range includes zero, so the detailed halo gas template remains weakly constrained by the present data.

{To test whether $\mathrm{DM}_{\rm model}$ tracks the variation in DM excess across the Pass sample, we add to the joint offset model a halo-template term centered on the Pass-sample mean, $\widetilde{H}_i=P_i\left(H_i-\langle H\rangle_{\rm Pass}\right)$:}
\begin{equation}
\mu_i
=
\alpha z_i
+
\Delta_{\rm Pass}P_i
+
A_{\rm trend}\widetilde{H}_i ,
\end{equation}
{with variance}
\begin{equation}
\sigma_{{\rm tot},i}^{2}
=
\sigma_{{\rm cosmic},i}^{2}
+
A_{\rm trend}^{2}\sigma_{H,i}^{2}
+
\sigma_{\rm int}^{2}.
\end{equation}

{We retain the priors and sampler settings of the joint offset model, adopt a uniform prior over $-10<A_{\rm trend}<10$, and run the chains for 12,000 steps. The fit gives $\alpha=983.6^{+165.2}_{-162.0}~\mathrm{pc\,cm^{-3}}$, $\Delta_{\rm Pass}=143.7^{+55.5}_{-55.6}~\mathrm{pc\,cm^{-3}}$, $A_{\rm trend}=0.004^{+0.483}_{-0.466}$, and $\sigma_{\rm int}=119.3^{+18.2}_{-15.5}~\mathrm{pc\,cm^{-3}}$. The fitted $\Delta_{\rm Pass}$ and $\sigma_{\rm int}$ are consistent with those from the primary joint fit.}

{The fitted $A_{\rm trend}$ is consistent with zero, and the observed scatter in $\mathrm{DM}_{\rm excess}$ is substantially larger than predicted by the adopted template. Together, these results indicate that a single rescaling of the template cannot reproduce the observed $\mathrm{DM}_{\rm excess}$ distribution.}

\section{Discussion}\label{sec:Discussion}
\subsection{Mean DM excess and the foreground halo template}
\label{sec:interpretation}

{The joint offset fit indicates that group-intersecting sight lines have a higher mean $\mathrm{DM}_{\rm cosmic}$ than clean nonintersecting sight lines. The centered template fit finds no measurable dependence of the residual variation within the Pass sample on the adopted foreground-halo template.}

{The foreground-halo template depends on several model ingredients.} The overall scale of $\mathrm{DM}_{\rm model}$ is sensitive to the adopted gas normalization and radial profile \citep[e.g.,][]{Wu2023ApJ...945...87W}. Its variation among sight lines also depends on halo masses, catalog centers, and impact parameters. Variations in gas structure and feedback can further alter the halo DM profiles and the scatter among sight lines \citep[e.g.,][]{Medlock2024ApJ...967...32M,Ghirardini2019A&A...621A..41G}. {These model dependencies may contribute to the broad $A_{\rm halo}$ posterior and the absence of a measurable centered-template trend.}

{The constraint on $A_{\rm halo}$ is also sensitive to individual sight lines, as illustrated by FRB 20181112A.} Previous work identified this burst as a case in which the sight line intersects a foreground halo but shows little apparent additional DM from the intervening system \citep{Prochaska2019Sci...366..231P}. In our analysis, it has a large modeled foreground contribution, $\mathrm{DM}_{\rm model}=238.4\pm32.4~\mathrm{pc\,cm^{-3}}$, but no positive residual excess relative to the NoPass baseline, $\mathrm{DM}_{\rm excess}=-22.5\pm85.7~\mathrm{pc\,cm^{-3}}$. As a diagnostic, excluding this sight line gives $A_{\rm halo}=1.08^{+0.63}_{-0.60}$. 

{We therefore interpret the measured offset as a population-level association between foreground group environments and additional DM.} This excess may arise from gas bound to the identified halos \citep{Connor2022NatAs...6.1035C} and correlated gas in surrounding filaments and other large-scale structures \citep{Hussaini2025ApJ...993L..27H,Mo2025ApJ...995..183M}. Their relative contributions remain uncertain.

\subsection{Sensitivity to sample and modeling choices}
\label{sec:robustness}

{To test the stability of the inferred mean offset between Pass and NoPass sight lines, we repeat the joint offset fit across the sample and modeling choices considered here. We also examine how these choices affect the foreground-halo template amplitude. These tests include the common redshift range, a fixed host DM prior, alternative Milky Way disk models, foreground richness cuts, joint propagation of the FRB localization and foreground-system redshift uncertainties, and a secure localization criterion.} For tests that change only the sample definition or the foreground halo template, the $\mathrm{DM}_{\rm cosmic}$ posterior for each FRB is kept fixed. For tests that change the Milky Way disk model or host DM prescription, the affected $\mathrm{DM}_{\rm cosmic}$ posteriors are recomputed before refitting.

{Recent work reports tentative evidence for an increase in host-galaxy DM with redshift \citep{Mas2026arXiv260625214M}. Because the Pass sample is concentrated at higher redshift than the NoPass sample, host-DM evolution not captured by the adopted prescription could contribute to the inferred $\Delta_{\rm Pass}$. To reduce sensitivity to the different redshift coverage of the two samples, we repeat the fit using only FRBs within their common redshift interval.}

The richness tests examine the effect of low richness systems, whose catalog centers and impact parameters are less secure \citep{Wang2022ApJ...936..161W}. We first remove low richness associations from $\mathrm{DM}_{\rm model}$ while leaving the Pass and NoPass samples unchanged. We then restrict the Pass sample to sight lines with higher richness foreground associations and retain the same clean NoPass sample.

{We assess the combined effects of FRB localization and foreground-system redshift uncertainties on the halo-template fit using $K=20{,}000$ Monte Carlo realizations for each Pass sight line. For each realization, we draw an FRB-position offset from an isotropic two-dimensional Gaussian with per-coordinate standard deviation $\sigma_{\rm maj}$ and independently draw $z_{\rm grp}^{(k)}\sim\mathcal{N}(z_{\rm grp},\sigma_z^2)$ for each associated foreground system, using a redshift treatment inspired by \citet{Simha2021ApJ...921..134S}. If a redshift draw places a foreground system at or beyond the FRB redshift, its contribution is set to zero. Holding the spectroscopic FRB redshifts and foreground-system centers fixed, we recompute the angular-diameter distances, impact parameters, redshift-dependent mass conversions and halo radii, aperture inclusion, and total foreground contribution $H_i^{(k)}$. We then refit the halo-template model by directly marginalizing the likelihood over the resulting $H_i^{(k)}$ distributions, using the same parameter priors as in the primary template fit.}

\begin{deluxetable*}{lccccc}[ht]
\tabletypesize{\scriptsize}
\tablewidth{0pt}
\tablecaption{Sensitivity tests for the offset between Pass and NoPass sight lines and the foreground-halo-template amplitude.\label{tab:robustness_test}}
\tablehead{
\colhead{Test} &
\colhead{$N_{\rm Pass}/N_{\rm NoPass}$} &
\colhead{$z$ range} &
\colhead{$\Delta_{\rm Pass}$} &
\colhead{$A_{\rm halo}$} &
\colhead{$P(A_{\rm halo}>0)$} \\
\colhead{} &
\colhead{} &
\colhead{} &
\colhead{(pc cm$^{-3}$)} &
\colhead{} &
\colhead{}
}
\startdata
Main analysis & 23/18 & $0.0008$--$0.4775$ & $143.3^{+53.2}_{-53.0}$ & $0.34^{+0.51}_{-0.50}$ & $0.751$ \\
Common redshift range\tablenotemark{a} & 12/10 & $0.0939$--$0.3212$ & $186.4^{+47.2}_{-48.7}$ & $1.04^{+0.73}_{-0.71}$ & $0.926$ \\
Fixed host-DM prior\tablenotemark{b} & 23/18 & $0.0008$--$0.4775$ & $110.9^{+54.2}_{-53.0}$ & $0.33^{+0.49}_{-0.47}$ & $0.753$ \\
NE2001 Milky Way disk model & 23/18 & $0.0008$--$0.4775$ & $144.5^{+52.7}_{-52.4}$ & $0.34^{+0.51}_{-0.49}$ & $0.751$ \\
YMW16 Milky Way disk model & 23/18 & $0.0008$--$0.4775$ & $142.7^{+53.6}_{-53.4}$ & $0.32^{+0.51}_{-0.49}$ & $0.738$ \\
Remove $R<3$ associations\tablenotemark{c} & 23/18 & $0.0008$--$0.4775$ & $143.3^{+53.2}_{-53.0}$ & $0.41^{+0.57}_{-0.56}$ & $0.766$ \\
Remove $R<4$ associations\tablenotemark{c} & 23/18 & $0.0008$--$0.4775$ & $143.3^{+53.2}_{-53.0}$ & $0.42^{+0.57}_{-0.57}$ & $0.771$ \\
Reliable $R\ge3$ associations\tablenotemark{d} & 16/18 & $0.0008$--$0.4775$ & $194.1^{+60.1}_{-61.0}$ & $0.25^{+0.64}_{-0.63}$ & $0.654$ \\
Reliable $R\ge4$ associations\tablenotemark{d} & 14/18 & $0.0008$--$0.4755$ & $201.8^{+59.5}_{-60.9}$ & $0.45^{+0.69}_{-0.69}$ & $0.742$ \\
Secure localization\tablenotemark{e} & 23/18 & $0.0008$--$0.4775$ & $143.3^{+53.2}_{-53.0}$ & $0.43^{+0.57}_{-0.57}$ & $0.772$ \\
Localization + foreground-$z$ propagation\tablenotemark{f} & 23/18 & $0.0008$--$0.4775$ & $143.3^{+53.2}_{-53.0}$ & $0.36^{+0.53}_{-0.50}$ & $0.764$ \\
\enddata
\tablenotetext{a}{
This test retains only FRBs within the common redshift interval,
$\max(z_{\rm min,Pass},z_{\rm min,NoPass})
\le z \le
\min(z_{\rm max,Pass},z_{\rm max,NoPass})$.
}
\tablenotetext{b}{
Following M25, this test uses a fixed lognormal prior for the rest-frame
host DM, with median $130~\mathrm{pc\,cm^{-3}}$ and logarithmic width
$\sigma_{\rm host}=0.56$.
}
\tablenotetext{c}{
Low-richness associations are removed only when calculating
$\mathrm{DM}_{\rm model}$; the Pass and NoPass samples are unchanged.
}
\tablenotetext{d}{
Pass sight lines without an association satisfying the stated richness
threshold are excluded; the NoPass sample is unchanged.
}
\tablenotetext{e}{
For this test, $\mathrm{DM}_{\rm model}$ is recalculated using only
$R>1$ associations satisfying $b+3\sigma_\perp\le r_{180}$, where
$\sigma_\perp=D_A(z_{\rm grp})\sigma_{\rm loc}$.
All $R=1$ contributions are excluded.
All 23 fiducial Pass sight lines retain at least one qualifying
$R>1$ association, so the sample sizes remain 23 and 18.
}
\tablenotetext{f}{
The sample labels and $\mathrm{DM}_{\rm cosmic}$ posteriors are unchanged
in this test, so $\Delta_{\rm Pass}$ is identical to the main analysis.
}
\tablecomments{
$A_{\rm halo}$ is dimensionless.
$\Delta_{\rm Pass}$ and $A_{\rm halo}$ are obtained from separate fits.
Quoted intervals are 16th--84th posterior percentiles.
}
\end{deluxetable*}

Table~\ref{tab:robustness_test} shows that the posterior median of $\Delta_{\rm Pass}$ remains positive across these sensitivity tests. The fixed host DM prior lowers the median from $143.3$ to $110.9~\mathrm{pc\,cm^{-3}}$ while leaving the $A_{\rm halo}$ posterior nearly unchanged, and alternative Milky Way disk models have little effect on either parameter. Restricting the redshift range or requiring richer foreground systems yields higher posterior medians for $\Delta_{\rm Pass}$, with uncertainty ranges that overlap the main result. The jackknife analysis also gives a positive median after each FRB is omitted in turn, with values spanning $108.3$--$159.1~\mathrm{pc\,cm^{-3}}$. The $A_{\rm halo}$ posteriors from the richness and localization tests remain broad. The common redshift range sample gives a higher median $A_{\rm halo}$; this sample excludes FRB 20181112A, whose influence is discussed in Section~\ref{sec:interpretation}. Overall, the foreground halo amplitude remains less well determined than the average offset.

We also perform random label null tests of the Pass and NoPass classification while keeping the observed redshifts and $\mathrm{DM}_{\rm cosmic}$ values fixed. We consider two relabeling schemes. In the unrestricted test, the labels are shuffled while preserving the total Pass and NoPass sample sizes. In the redshift binned tests, the combined sample is sorted by redshift and divided into three to six bins containing approximately equal numbers of FRBs. The labels are then shuffled within each bin while preserving its Pass and NoPass counts. For each relabeled sample, we refit the joint offset model by profile maximum likelihood and use $T=\widehat{\Delta}_{\rm Pass}$ as the test statistic. Using $N_{\rm perm}=100{,}000$ relabelings for each setup, we estimate the upper tail probability as $\widehat{p}=(k+1)/(N_{\rm perm}+1)$, where $k$ is the number of relabelings with $T\ge T_{\rm obs}$. For the main sample, the unrestricted test gives $\widehat{p}=6.0\times10^{-5}$, and the redshift binned tests give $\widehat{p}=5.0\times10^{-5}$--$8.8\times10^{-4}$. The corresponding values for the common redshift range sample are $1.0\times10^{-3}$ and $1.6\times10^{-3}$--$4.5\times10^{-3}$. Thus, fitted offsets equal to or greater than the observed value are uncommon under both relabeling schemes.

\subsection{Forecast for the foreground-halo template constraint}

The broad posterior distribution of $A_{\rm halo}$ motivates a forecast of how its uncertainty scales with additional localized FRBs. {We add mock FRBs with sky positions drawn uniformly in solid angle within the Y21/DESI group-catalog footprint. Their redshifts are drawn from the detected-FRB redshift distributions predicted for the Deep Synoptic Array (DSA), CRAFT/ASKAP, and MeerTRAP using zDM, a framework for modeling the FRB redshift--dispersion-measure distribution \citep{James2022MNRAS.509.4775J}. We treat the three distributions as separate forecast scenarios and restrict all draws to the primary-analysis range, $0<z<0.5$.}

{Using the same Y21 matching and halo-template procedure as for the observed sample, we classify each mock sight line as Pass or NoPass and calculate $H_i=\mathrm{DM}_{{\rm model},i}$ and $\sigma_{H,i}$ for Pass sight lines. For NoPass sight lines, $H_i=\sigma_{H,i}=0$. The combined number of Pass and NoPass sight lines defines $N_{\rm add}$.}

{For each realization, we draw $(\alpha_{\rm h}^{\rm inj},A_{\rm halo}^{\rm inj},\sigma_{\rm int,h}^{\rm inj})$ from the joint posterior of the observed foreground-halo template fit and generate}
\begin{equation}
\mathrm{DM}_{{\rm cosmic},i}^{\rm mock}
=
\alpha_{\rm h}^{\rm inj}z_i
+
A_{\rm halo}^{\rm inj}H_i
+
\epsilon_i .
\end{equation}
{Here, $\epsilon_i$ is a zero-mean Gaussian residual with variance}
\begin{equation}
\sigma_{\epsilon,i}^2
=
\sigma_{{\rm cosmic},i}^2
+
\left(A_{\rm halo}^{\rm inj}\right)^2\sigma_{H,i}^2
+
\left(\sigma_{\rm int,h}^{\rm inj}\right)^2 .
\end{equation}

{The uncertainty $\sigma_{{\rm cosmic},i}$ for each mock FRB is assigned by randomly drawing one of the symmetrized uncertainties of the eight FRBs in the observed fitting sample closest in redshift. Each realization is analyzed using the same procedure as the observed halo-template fit.} We define $\sigma(A_{\rm halo})$ as half the difference between the 84th and 16th percentiles of the posterior.

\begin{figure}[t]
\centering
\includegraphics[width=\linewidth]{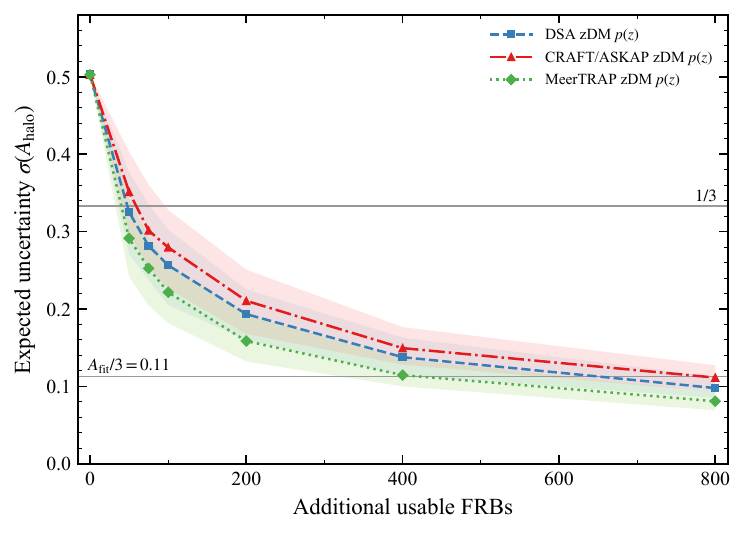}
\caption{{Forecasted uncertainty in $A_{\rm halo}$ as a function of the number of additional usable FRBs. The three curves correspond to the zDM redshift distributions for DSA, CRAFT/ASKAP, and MeerTRAP. The curves and shaded regions show the median and 16th--84th percentile ranges across 300 mock realizations. The horizontal lines mark the precision benchmarks $\sigma(A_{\rm halo})=1/3$ and $\sigma(A_{\rm halo})=A_{\rm halo}^{\rm fit}/3=0.11$.}}
\label{fig:forecast}
\end{figure}

Figure~\ref{fig:forecast} shows that the expected uncertainty decreases as additional usable FRBs are added. {Across the three zDM redshift distributions, the median forecast reaches the fitted-amplitude reference, $\sigma(A_{\rm halo})=A_{\rm halo}^{\rm fit}/3=0.11$ for $A_{\rm halo}^{\rm fit}=0.34$, after approximately 420--780 additional usable FRBs. Because $A_{\rm halo}^{\rm fit}$ is sensitive to individual sight lines (see Section~\ref{sec:interpretation}), we also report $\sigma(A_{\rm halo})=1/3$ as a reference not tied to the fitted value. This level is reached after approximately 40--60 additional usable FRBs and would correspond to a nominal $3\sigma$ constraint if the true amplitude were $A_{\rm halo}=1$.}

\section{Conclusions}\label{sec:Conclusions}

In this work, we construct a parent sample of 117 localized FRBs with measured DMs and host redshifts, and use the Y21/DESI group catalog to identify foreground halo intersections along their sight lines. For the main analysis, we restrict the sample to FRBs within the group catalog footprint and with $z_{\rm host}<0.5$, where the foreground catalog provides a more uniform basis for the comparison. This gives 23 sight lines intersecting foreground groups and 18 clean nonintersecting sight lines. Our main results are as follows:

\begin{enumerate}

    \item Using a joint fit with a common $\mathrm{DM}_{\rm cosmic}$--$z$ relation and an additional offset for Pass sight lines, we find $\Delta_{\rm Pass}=143.3^{+53.2}_{-53.0}~\mathrm{pc\,cm^{-3}}$. Thus, sight lines intersecting foreground groups have higher $\mathrm{DM}_{\rm cosmic}$ on average than clean nonintersecting sight lines. Random label tests that preserve the Pass and NoPass counts within redshift bins give estimated one-sided probabilities of $5.0\times10^{-5}$--$8.8\times10^{-4}$ for three to six bins. {Complementary regression and distributional comparisons also support this difference, with the inferred statistical strength varying among methods.}

    \item {The foreground-halo template amplitude is weakly constrained, with $A_{\rm halo}=0.34^{+0.51}_{-0.50}$ and $P(A_{\rm halo}>0)=0.751$. The centered joint fit separately accounts for the mean Pass-sample offset and tests whether the relative variation predicted by $\mathrm{DM}_{\rm model}$ follows the remaining variation in $\mathrm{DM}_{\rm cosmic}$. It gives a mean offset consistent with the primary result and $A_{\rm trend}=0.004^{+0.483}_{-0.466}$, showing no measurable correspondence between the two variations.}

    \item The DESI footprint mock forecast shows that larger localized FRB samples would improve the constraint on $A_{\rm halo}$. {Using the current fitted median, $A_{\rm halo}^{\rm fit}=0.34$, as the precision reference, the median expected uncertainty reaches $\sigma(A_{\rm halo})=A_{\rm halo}^{\rm fit}/3=0.11$ with approximately 420--780 additional usable FRBs across the three zDM redshift distributions.}

\end{enumerate}

Future progress will require larger samples of well-localized FRBs and improved foreground catalogs. Better constraints on Milky Way, host galaxy, foreground catalog, and foreground halo modeling uncertainties will reduce the effective scatter in the template fit and enable more direct tests of baryons in foreground galaxy groups and clusters with FRB dispersion measures.

\begin{acknowledgments}
We thank Weishan Zhu for his helpful discussion. This work is supported by the National SKA Program of China No. 2025SKA0150103, and National Natural Science Foundation of China under Nos. 12550002, 12133008, 12221003, and 11890692. We acknowledge the science research grants from the China Manned Space Project with No. CMS-CSST-2021-A04 and No. CMS-CSST-2025-A10.
\end{acknowledgments}

\bibliography{sample701}{}
\bibliographystyle{aasjournalv7}

\clearpage
\onecolumngrid
\appendix
\setcounter{figure}{0}
\renewcommand{\thefigure}{A\arabic{figure}}
\setcounter{table}{0}
\renewcommand{\thetable}{A\arabic{table}}
\renewcommand{\theHtable}{A\arabic{table}}

\section{Localized Fast Radio Burst Catalog}  

Table~\ref{tab:frb-sample} lists the adopted properties and localization references for the 117 FRBs in the parent sample.

\startlongtable
\begin{deluxetable}{lccccccccc}
\tabletypesize{\scriptsize}
\tablecaption{Verified Localized Fast Radio Burst Catalog\label{tab:frb-sample}}
\tablehead{
\colhead{FRB ID} &
\colhead{R.A.} &
\colhead{Decl.} &
\colhead{$\sigma_{\rm maj}$} &
\colhead{DM} &
\colhead{DM$_{\rm MW,disk}$\tablenotemark{a}} &
\colhead{DM$_{\rm MW,halo}$\tablenotemark{b}} &
\colhead{$z_{\rm host}$} &
\colhead{$\sigma_{\rm cosmic}$} &
\colhead{Loc. References} \\
\colhead{} &
\colhead{(deg)} &
\colhead{(deg)} &
\colhead{(arcsec)} &
\colhead{(pc cm$^{-3}$)} &
\colhead{(pc cm$^{-3}$)} &
\colhead{(pc cm$^{-3}$)} &
\colhead{} &
\colhead{(pc cm$^{-3}$)} &
\colhead{}
}
\startdata
20121102A & 82.994589 & 33.147932 & 0.004 & 558.10 & 207.93 & 40.91 & 0.19273 & 111.68 & (1) \\
20180301A & 93.226833 & 4.671056 & 0.619 & 536.00 & 156.17 & 38.06 & 0.33044 & 96.24 & (1) \\
20180916B & 29.503126 & 65.716754 & 0.002 & 348.76 & 125.65 & 43.08 & 0.03370 & 63.16 & (1) \\
20180924B & 326.105229 & -40.900028 & 0.176 & 362.16 & 39.79 & 45.54 & 0.32120 & 53.37 & (1) \\
20181112A & 327.348458 & -52.970944 & 5.755 & 589.27 & 39.47 & 45.04 & 0.47550 & 44.12 & (1) \\
20181119A & 190.479946 & 65.111920 & 0.010 & 366.10 & 31.20 & 31.35 & 0.26064 & 60.05 & (1) \\
20190102C & 322.415667 & -79.475694 & 0.959 & 364.55 & 55.04 & 46.99 & 0.29120 & 62.57 & (1) \\
20190417A & 294.774550 & 59.326897 & 0.005 & 1378.90 & 76.20 & 43.27 & 0.12817 & 80.56 & (2) \\
20190425A & 255.675000 & 21.576389 & 12.000 & 128.20 & 48.35 & 48.98 & 0.03122 & 17.48 & (3) \\
20190520B & 240.517775 & -11.288147 & 0.100 & 1210.30 & 60.37 & 69.16 & 0.24100 & 76.91 & (1) \\
20190523A & 207.065000 & 72.469722 & 4.000 & 760.80 & 34.99 & 32.39 & 0.66000 & 52.21 & (1) \\
20190608B & 334.019875 & -7.898250 & 0.381 & 340.05 & 36.10 & 38.89 & 0.11778 & 57.80 & (1) \\
20190611B & 320.745583 & -79.397583 & 1.170 & 332.63 & 55.44 & 47.21 & 0.37780 & 58.32 & (1) \\
20190711A & 329.417225 & -80.358029 & 2.409 & 587.90 & 54.05 & 46.20 & 0.52172 & 61.53 & (1) \\
20190714A & 183.979708 & -13.021000 & 0.546 & 504.13 & 35.73 & 36.44 & 0.23650 & 65.63 & (1) \\
20191001A & 323.351696 & -54.748306 & 0.460 & 507.00 & 41.84 & 46.71 & 0.23400 & 63.42 & (1) \\
20191228A & 344.430529 & -29.594117 & 0.840 & 298.00 & 30.23 & 36.88 & 0.24300 & 54.62 & (1) \\
20200120E & 149.477833 & 68.818889 & 0.400 & 87.82 & 36.35 & 31.10 & 0.00080 & 10.29 & (1) \\
20200430A & 229.706454 & 12.376329 & 0.459 & 380.00 & 29.54 & 42.24 & 0.16100 & 38.50 & (1) \\
20200906A & 53.495546 & -14.083008 & 0.510 & 577.84 & 32.66 & 30.16 & 0.36880 & 48.83 & (1) \\
20201124A & 77.014615 & 26.060696 & 0.004 & 411.00 & 136.07 & 36.24 & 0.09820 & 78.77 & (1) \\
20210117A & 339.979196 & -16.151445 & 0.418 & 728.95 & 32.26 & 37.17 & 0.21450 & 67.06 & (1) \\
20210320C & 204.458767 & -16.122670 & 0.471 & 384.59 & 37.86 & 42.13 & 0.27970 & 61.97 & (1) \\
20210405I & 255.339700 & -49.545150 & 0.500 & 565.17 & 444.25 & 143.95 & 0.06600 & 115.16 & (4) \\
20210410D & 326.086250 & -79.318194 & 0.800 & 575.00 & 53.88 & 46.48 & 0.14150 & 74.69 & (1) \\
20210603A & 10.274058 & 21.226270 & 1.080 & 500.15 & 36.90 & 32.43 & 0.17720 & 66.16 & (5) \\
20210807D & 299.221433 & -0.762361 & 0.500 & 251.30 & 117.05 & 73.58 & 0.12927 & 39.01 & (1) \\
20211127I & 199.808812 & -18.837911 & 0.510 & 234.97 & 40.27 & 41.59 & 0.04690 & 40.53 & (1) \\
20211203C & 204.562500 & -31.380267 & 0.440 & 635.00 & 60.52 & 49.79 & 0.34386 & 69.00 & (1) \\
20211212A & 157.350792 & 1.360431 & 0.500 & 209.00 & 34.72 & 32.18 & 0.07070 & 37.86 & (1) \\
20220105A & 208.803354 & 22.466228 & 1.513 & 580.00 & 23.39 & 34.35 & 0.27850 & 44.63 & (1) \\
20220204A & 274.226250 & 69.722503 & 1.560 & 612.20 & 50.01 & 37.00 & 0.40120 & 44.03 & (1) \\
20220207C & 310.199525 & 72.882327 & 0.630 & 262.30 & 71.00 & 39.21 & 0.04330 & 44.66 & (1) \\
20220208A & 322.575125 & 70.041042 & 0.990 & 437.00 & 93.55 & 41.46 & 0.35100 & 73.04 & (1) \\
20220222C & 203.904500 & -28.026925 & 0.550 & 1071.20 & 53.19 & 47.40 & 0.85300 & 54.75 & (6) \\
20220224C & 166.677542 & -22.939856 & 0.640 & 1140.20 & 47.81 & 37.67 & 0.62710 & 55.88 & (6) \\
20220307B & 350.874496 & 72.192386 & 1.720 & 499.15 & 111.79 & 41.27 & 0.24810 & 82.81 & (1) \\
20220310F & 134.720500 & 73.490833 & 1.900 & 462.15 & 41.18 & 31.81 & 0.47800 & 42.79 & (1) \\
20220319D & 32.177921 & 71.035260 & 0.580 & 110.95 & 99.41 & 39.69 & 0.01120 & 18.41 & (1) \\
20220330D & 163.751250 & 70.350750 & 1.620 & 468.10 & 34.92 & 31.22 & 0.37140 & 51.36 & (1) \\
20220418A & 219.105575 & 70.095944 & 1.356 & 623.45 & 34.91 & 32.84 & 0.62130 & 42.59 & (1) \\
20220501C & 352.379113 & -32.490731 & 0.440 & 449.50 & 28.26 & 35.26 & 0.38100 & 58.66 & (1) \\
20220506D & 318.044833 & 72.827278 & 1.110 & 396.93 & 78.14 & 39.75 & 0.30040 & 67.91 & (1) \\
20220509G & 282.670000 & 70.243833 & 2.200 & 269.50 & 54.37 & 37.70 & 0.08940 & 48.84 & (1) \\
20220529A & 19.104125 & 20.632375 & 1.390 & 246.40 & 36.55 & 31.65 & 0.18390 & 37.85 & (7) \\
20220610A & 351.073262 & -33.513850 & 0.445 & 1458.10 & 28.58 & 35.64 & 1.01500 & 44.15 & (1) \\
20220717A & 293.304167 & -19.287722 & 0.900 & 634.69 & 106.08 & 86.47 & 0.36295 & 88.93 & (8) \\
20220725A & 353.315204 & -35.990256 & 0.460 & 290.40 & 28.54 & 35.42 & 0.19260 & 54.24 & (1) \\
20220726A & 73.945542 & 69.929528 & 0.800 & 686.55 & 81.18 & 35.32 & 0.36190 & 71.65 & (1) \\
20220825A & 311.981450 & 72.584969 & 0.780 & 651.20 & 73.12 & 39.45 & 0.24140 & 74.11 & (1) \\
20220831A & 338.695542 & 70.538444 & 0.480\tablenotemark{c} & 1146.25 & 118.24 & 42.10 & 0.26200 & 87.93 & (9) \\
20220912A & 347.270412 & 48.706641 & 0.005 & 219.37 & 115.68 & 43.44 & 0.07710 & 37.98 & (10) \\
20220914A & 282.056792 & 73.336914 & 0.930 & 631.05 & 52.95 & 36.95 & 0.11390 & 43.26 & (1) \\
20220918A & 17.592142 & -70.811378 & 0.460 & 656.80 & 39.04 & 38.66 & 0.49100 & 57.89 & (1) \\
20220920A & 240.257083 & 70.918800 & 1.020 & 315.00 & 38.70 & 34.12 & 0.15820 & 46.84 & (1) \\
20221012A & 280.798708 & 70.524208 & 1.120 & 442.20 & 53.22 & 37.46 & 0.28470 & 64.73 & (1) \\
20221027A & 130.872000 & 72.100944 & 0.800 & 452.50 & 41.74 & 31.76 & 0.54220 & 35.94 & (1) \\
20221029A & 141.963417 & 72.452317 & 0.800 & 1391.05 & 39.18 & 31.56 & 0.97500 & 41.69 & (1) \\
20221101B & 342.215792 & 70.681500 & 1.200 & 490.70 & 123.10 & 42.13 & 0.23950 & 85.62 & (1) \\
20221106A & 56.704808 & -25.569810 & 0.598 & 343.80 & 32.58 & 30.71 & 0.20440 & 60.11 & (1) \\
20221113A & 71.411000 & 70.307389 & 1.400 & 411.40 & 83.44 & 35.54 & 0.25050 & 70.65 & (1) \\
20221116A & 21.210958 & 72.653750 & 1.000 & 640.60 & 99.53 & 39.85 & 0.27640 & 80.57 & (1) \\
20221219A & 257.629792 & 71.626842 & 1.540 & 706.71 & 43.46 & 35.32 & 0.55300 & 45.69 & (1) \\
20230124A & 231.917000 & 70.968086 & 1.400 & 590.00 & 37.11 & 33.58 & 0.09390 & 42.06 & (1) \\
20230125D & 150.205042 & -31.544658 & 0.260 & 640.08 & 76.85 & 41.76 & 0.32650 & 72.39 & (6) \\
20230203A & 151.661594 & 35.694096 & 17.978 & 417.30 & 31.01 & 29.90 & 0.14640 & 66.19 & (1) \\
20230216A & 156.472167 & 3.436825 & 1.200 & 828.29 & 34.30 & 31.91 & 0.53100 & 38.19 & (1) \\
20230222B & 238.739123 & 30.898695 & 11.498 & 188.03 & 29.90 & 38.96 & 0.11000 & 30.73 & (1) \\
20230307A & 177.781583 & 71.695131 & 0.200 & 610.15 & 34.44 & 31.51 & 0.27060 & 59.25 & (1) \\
20230311A & 91.109662 & 55.945953 & 15.739 & 364.34 & 82.93 & 34.27 & 0.19180 & 66.53 & (1) \\
20230501A & 340.027083 & 70.922139 & 0.900 & 533.70 & 117.19 & 41.90 & 0.30150 & 84.50 & (1) \\
20230506C & 12.099837 & 42.006090 & 0.180 & 772.00 & 63.64 & 35.96 & 0.38960 & 65.97 & (11) \\
20230521B & 351.036000 & 71.138031 & 0.559\tablenotemark{c} & 1342.90 & 120.81 & 41.95 & 1.35400 & 72.58 & (9) \\
20230526A & 22.232617 & -52.717331 & 0.430 & 361.40 & 30.37 & 34.06 & 0.15700 & 62.81 & (1) \\
20230613A & 356.852708 & -27.052781 & 0.520 & 483.51 & 27.56 & 33.95 & 0.39230 & 58.31 & (6) \\
20230616A & 184.614579 & 28.356649 & 16.992 & 918.33 & 21.55 & 31.13 & 0.44850 & 55.61 & (1) \\
20230626A & 235.629583 & 71.133547 & 1.800 & 451.20 & 37.83 & 33.81 & 0.32700 & 52.56 & (1) \\
20230628A & 166.786708 & 72.281844 & 1.400 & 345.15 & 35.51 & 31.44 & 0.12700 & 37.52 & (1) \\
20230702A & 132.124693 & 26.389497 & 15.264 & 598.88 & 41.54 & 30.81 & 0.36580 & 61.24 & (1) \\
20230708A & 303.115542 & -55.356270 & 0.467 & 411.51 & 57.77 & 58.03 & 0.10500 & 71.42 & (1) \\
20230712A & 167.358529 & 72.557783 & 1.400 & 586.96 & 35.61 & 31.47 & 0.45250 & 59.02 & (1) \\
20230718A & 128.161929 & -40.451932 & 0.610 & 477.00 & 490.24 & 56.40 & 0.03500 & 114.36 & (1) \\
20230730A & 54.664564 & 33.159296 & 12.215 & 312.58 & 75.33 & 33.75 & 0.21150 & 57.63 & (1) \\
20230808F & 53.304125 & -51.935283 & 0.500 & 653.20 & 34.11 & 33.38 & 0.34720 & 61.11 & (12) \\
20230814B & 335.974750 & 73.025906 & 0.792\tablenotemark{c} & 696.40 & 96.64 & 40.53 & 0.55350 & 72.51 & (9) \\
20230828A & 22.287034 & 34.429645 & 12.485 & 377.28 & 49.70 & 33.39 & 0.62680 & 54.43 & (1) \\
20230902A & 52.139775 & -47.333503 & 0.693 & 440.10 & 32.82 & 32.63 & 0.36190 & 59.35 & (1) \\
20230907D & 187.142500 & 8.658092 & 0.570 & 1030.79 & 26.88 & 33.02 & 0.46380 & 55.98 & (6) \\
20230918A & 161.050726 & 61.120861 & 11.617 & 365.96 & 31.98 & 30.48 & 0.06000 & 65.94 & (1) \\
20230923A & 206.614104 & 45.408008 & 13.813 & 918.33 & 26.50 & 32.17 & 0.44450 & 56.44 & (1) \\
20230926A & 269.124876 & 41.814304 & 17.964 & 222.81 & 55.57 & 44.22 & 0.05530 & 35.40 & (1) \\
20230930A & 10.507225 & 41.417337 & 0.180 & 456.00 & 62.16 & 35.96 & 0.09250 & 73.87 & (11) \\
20231005A & 246.028004 & 35.448712 & 11.390 & 188.84 & 34.29 & 39.60 & 0.07130 & 29.90 & (1) \\
20231006B & 109.619968 & 84.776136 & 16.042 & 1252.30 & 49.60 & 33.67 & 0.47800 & 59.56 & (1) \\
20231011A & 18.241101 & 41.749098 & 11.495 & 186.42 & 63.55 & 35.36 & 0.07830 & 27.61 & (1) \\
20231017A & 346.754287 & 36.652682 & 13.122 & 344.12 & 62.44 & 39.10 & 0.24500 & 61.43 & (1) \\
20231020B & 57.278208 & -37.769878 & 0.400 & 952.20 & 32.94 & 31.66 & 0.47750 & 56.17 & (6) \\
20231102A & 228.837416 & 56.799192 & 15.008 & 906.20 & 32.41 & 33.76 & 0.68780 & 50.71 & (1) \\
20231120A & 143.983958 & 73.284667 & 1.100 & 438.90 & 39.14 & 31.62 & 0.03680 & 61.92 & (1) \\
20231123A & 82.623249 & 4.475541 & 16.254 & 302.07 & 79.56 & 34.22 & 0.07290 & 54.87 & (1) \\
20231123B & 242.538167 & 70.785056 & 1.200 & 396.70 & 39.14 & 34.28 & 0.26210 & 62.38 & (1) \\
20231220A & 123.908708 & 73.659917 & 0.326\tablenotemark{c} & 491.20 & 44.16 & 32.12 & 0.33550 & 62.47 & (9) \\
20231223C & 259.544649 & 29.497937 & 11.333 & 179.54 & 48.83 & 46.34 & 0.10590 & 25.90 & (1) \\
20231226A & 155.363733 & 6.110260 & 0.511 & 329.90 & 33.82 & 31.60 & 0.15690 & 59.87 & (1) \\
20231229A & 26.467825 & 35.112918 & 11.430 & 198.55 & 51.85 & 33.35 & 0.01900 & 30.78 & (1) \\
20240114A & 321.915979 & 4.329352 & 0.093 & 527.65 & 51.05 & 45.05 & 0.13029 & 73.84 & (13) \\
20240123A & 68.262500 & 71.945283 & 0.606\tablenotemark{c} & 1462.00 & 82.11 & 35.65 & 0.96800 & 59.87 & (9) \\
20240201A & 149.905608 & 14.087972 & 0.457 & 374.50 & 33.78 & 30.88 & 0.04273 & 67.28 & (1) \\
20240209A & 289.887500 & 86.064444 & 2.120 & 176.49 & 51.43 & 34.92 & 0.13840 & 26.29 & (14) \\
20240210A & 8.779575 & -28.270754 & 0.547 & 283.73 & 26.38 & 32.38 & 0.02369 & 54.91 & (1) \\
20240213A & 166.168292 & 74.075389 & 0.466\tablenotemark{c} & 357.40 & 36.49 & 31.63 & 0.11850 & 63.60 & (9) \\
20240215A & 268.441250 & 70.232361 & 0.373\tablenotemark{c} & 549.50 & 47.31 & 36.38 & 0.21000 & 68.89 & (9) \\
20240304A & 136.330825 & -16.166631 & 0.810 & 652.60 & 65.09 & 37.09 & 0.24230 & 71.77 & (1) \\
20240304B & 182.997042 & 11.813017 & 0.480 & 2458.20 & 26.27 & 32.26 & 2.14800 & 32.17 & (15) \\
20240310A & 17.621871 & -44.439424 & 0.588 & 601.80 & 28.29 & 33.15 & 0.12700 & 70.35 & (1) \\
20241228A & 216.385750 & 12.025042 & 10.930 & 246.53 & 25.65 & 37.99 & 0.16140 & 46.39 & (16) \\
20250316A & 182.434663 & 58.849086 & 0.068 & 161.82 & 29.26 & 30.78 & 0.00650 & 24.32 & (17) \\
\enddata
\tablecomments{
The table lists FRBs with spectroscopic host-galaxy redshifts and adopted localization semimajor axis $\sigma_{\rm maj}<20''$. Here $\sigma_{\rm maj}=\max(\sigma_1,\sigma_2)$, where $\sigma_1$ and $\sigma_2$ are the adopted 1$\sigma$ uncertainties along the two localization axes. The final column gives the source of the adopted FRB position and localization uncertainty.
}
\tablerefs{
(1) \citet{Prochaska2025zndo..16763432P}\tablenotemark{d};
(2) \citet{Moroianu2026ApJ...996L..16M};
(3) \citet{Bhardwaj2024ApJ...971L..51B};
(4) \citet{Driessen2024MNRAS.527.3659D};
(5) \citet{Cassanelli2024NatAs...8.1429C};
(6) \citet{Pastor2026MNRAS.545f2144P};
(7) \citet{Pandhi2026ApJ..1000L..53P};
(8) \citet{Rajwade2024MNRAS.532.3881R};
(9) \citet{Connor2025NatAs...9.1226C};
(10) \citet{Hewitt2024MNRAS.529.1814H};
(11) \citet{Anna-Thomas2025ApJ...993..221A};
(12) \citet{Hanmer2025MNRAS.538.1800H};
(13) \citet{Bhardwaj2025ApJ...992L..35B};
(14) \citet{Shah2025ApJ...979L..21S};
(15) \citet{Caleb2025arXiv250801648C};
(16) \citet{Curtin2026ApJ...998...97C};
(17) \citet{CHIME2025ApJ...989L..48C}.
}
\tablenotetext{a}{
Fiducial Milky Way disk contribution estimated with the NE2025 Galactic
disk electron-density model \citep{Ocker2026ApJ..1002....3O}.
}
\tablenotetext{b}{
Fiducial Milky Way halo contribution estimated with the YT2020 Milky Way
halo model \citep{Yamasaki2020ApJ...888..105Y}.
}
\tablenotetext{c}{
For the DSA-110 rows marked in the $\sigma_{\rm maj}$ column, the
localization axes are from \citet{Connor2025NatAs...9.1226C}.
The DSA-110 localization procedure and 90\% confidence convention are
described by \citet{Sharma2024Natur.635...61S}; these axes are converted
to the adopted 1$\sigma$ axes before taking $\sigma_{\rm maj}$.
}
\tablenotetext{d}{
This Zenodo release corresponds to the public FRB
compilation; the associated database repository is available at
\url{https://github.com/FRBs/FRB}.
}
\end{deluxetable}

\onecolumngrid

\setcounter{figure}{0}
\renewcommand{\thefigure}{B\arabic{figure}}
\setcounter{table}{0}
\renewcommand{\thetable}{B\arabic{table}}
\renewcommand{\theHtable}{B\arabic{table}}

\section{Fast Radio Burst--Group Intersection Catalog}  Table~\ref{tab:frb_groups} lists the 61 FRB sight lines within the Y21/DESI group-catalog footprint, their classifications, and the identified foreground systems.
\begin{longrotatetable}
\begin{deluxetable}{ccccccccccccc}
\tabletypesize{\scriptsize}
\tablecaption{FRB sight lines within the Y21/DESI group-catalog footprint and their identified foreground systems.\label{tab:frb_groups}}
\tablehead{
\colhead{Object ID} &
\colhead{$\mathrm{RA}_{\rm FRB}$} &
\colhead{$\mathrm{Dec}_{\rm FRB}$} &
\colhead{$z_{\rm FRB}$} &
\colhead{DM$_{\rm obs}$} &
\colhead{Group ID\tablenotemark{1}} &
\colhead{Richness} &
\colhead{$\log_{10} M_h$\tablenotemark{2}} &
\colhead{$\mathrm{RA}_{\rm grp}$} &
\colhead{$\mathrm{Dec}_{\rm grp}$} &
\colhead{$b_\perp$\tablenotemark{3}} &
\colhead{$R_{\rm halo}$\tablenotemark{4}} &
\colhead{$z_{\rm grp}$} \\
\colhead{} &
\colhead{(deg)} &
\colhead{(deg)} &
\colhead{} &
\colhead{(pc cm$^{-3}$)} &
\colhead{} &
\colhead{} &
\colhead{} &
\colhead{(deg)} &
\colhead{(deg)} &
\colhead{(kpc)} &
\colhead{(kpc)} &
\colhead{}
}
\startdata
\cutinhead{group-intersecting sight lines ($R>1$)}
FRB 20181112A & 327.3485 & -52.9709 & 0.4755 & 589.27 & 3306298 & 2 & 12.1896 & 327.4587 & -52.9878 & 293.24 & 400.77 & 0.0595 \\
 &  &  &  &  & 31873928 & 1 & 12.2845 & 327.3530 & -52.9901 & 222.57 & 324.01 & 0.1843 \\
 &  &  &  &  & 31873874 & 1 & 12.1024 & 327.3664 & -52.9578 & 262.68 & 269.05 & 0.2717 \\
 &  &  &  &  & 54848 & 20 & 13.9568 & 327.3451 & -52.9347 & 600.49 & 1268.21 & 0.2998 \\
 &  &  &  &  & 662882 & 5 & 13.2658 & 327.3101 & -52.9960 & 605.49 & 727.46 & 0.3333 \\
 &  &  &  &  & 2414201 & 3 & 12.6778 & 327.3602 & -52.9550 & 315.59 & 460.03 & 0.3426 \\
 &  &  &  &  & 3306304 & 2 & 12.6374 & 327.3721 & -52.9837 & 367.17 & 435.10 & 0.3762 \\
 &  &  &  &  & 31873924 & 1 & 12.3402 & 327.3496 & -52.9696 & 29.25 & 304.55 & 0.3815 \\
 &  &  &  &  & 3306305 & 2 & 12.6402 & 327.3703 & -52.9651 & 297.96 & 420.92 & 0.4256 \\
\hline
FRB 20190523A & 207.0650 & 72.4697 & 0.6600 & 760.80 & 1037995 & 4 & 12.3607 & 207.0468 & 72.6371 & 414.10 & 468.61 & 0.0333 \\
 &  &  &  &  & 276700 & 9 & 13.0980 & 207.5319 & 72.6017 & 654.19 & 814.82 & 0.0465 \\
 &  &  &  &  & 10691 & 42 & 13.9777 & 206.6487 & 72.4219 & 1055.67 & 1500.02 & 0.1167 \\
 &  &  &  &  & 2933795 & 3 & 12.7635 & 207.0862 & 72.4699 & 114.96 & 492.34 & 0.3398 \\
 &  &  &  &  & 916429 & 5 & 13.5792 & 207.0540 & 72.4710 & 81.60 & 814.85 & 0.5140 \\
\hline
FRB 20191001A & 323.3517 & -54.7483 & 0.2340 & 507.00 & 31820861 & 1 & 11.0908 & 323.3523 & -54.7273 & 87.14 & 139.42 & 0.0575 \\
 &  &  &  &  & 31820976 & 1 & 10.9535 & 323.4007 & -54.7474 & 123.44 & 125.31 & 0.0607 \\
 &  &  &  &  & 282991 & 9 & 13.3416 & 323.5897 & -54.7712 & 896.76 & 940.46 & 0.0931 \\
\hline
FRB 20200430A & 229.7065 & 12.3763 & 0.1610 & 380.00 & 36190121 & 1 & 11.9461 & 229.6937 & 12.3678 & 66.86 & 267.32 & 0.0617 \\
 &  &  &  &  & 1076709 & 4 & 12.3691 & 229.6893 & 12.3651 & 151.49 & 438.73 & 0.1108 \\
\hline
FRB 20200906A & 53.4955 & -14.0830 & 0.3688 & 577.84 & 2105959 & 3 & 12.7071 & 53.4680 & -14.0372 & 326.78 & 580.27 & 0.0886 \\
 &  &  &  &  & 5766940 & 2 & 12.8325 & 53.5363 & -14.0778 & 444.81 & 590.71 & 0.1774 \\
 &  &  &  &  & 34647117 & 1 & 11.8825 & 53.4958 & -14.0832 & 5.23 & 223.45 & 0.3049 \\
\hline
FRB 20220105A & 208.8034 & 22.4662 & 0.2785 & 580.00 & 27545335 & 1 & 11.7781 & 208.8274 & 22.4312 & 189.50 & 234.88 & 0.0638 \\
 &  &  &  &  & 5619880 & 2 & 12.3018 & 208.7777 & 22.4355 & 361.83 & 405.05 & 0.1426 \\
 &  &  &  &  & 27545312 & 1 & 11.1886 & 208.8066 & 22.4777 & 136.92 & 140.18 & 0.1852 \\
\hline
FRB 20220204A & 274.2262 & 69.7225 & 0.4012 & 612.20 & 185 & 187 & 14.8214 & 273.6467 & 69.6517 & 1292.90 & 2944.68 & 0.0870 \\
\hline
FRB 20220418A & 219.1056 & 70.0959 & 0.6213 & 623.45 & 10647718 & 1 & 10.8863 & 219.0565 & 70.0781 & 35.64 & 121.73 & 0.0193 \\
 &  &  &  &  & 10647638 & 1 & 12.4573 & 219.0921 & 70.0936 & 71.17 & 360.09 & 0.2335 \\
 &  &  &  &  & 863881 & 5 & 13.0274 & 219.2023 & 70.1108 & 537.70 & 641.87 & 0.2584 \\
 &  &  &  &  & 10647637 & 1 & 11.9640 & 219.0941 & 70.0973 & 86.38 & 222.49 & 0.4323 \\
 &  &  &  &  & 10647571 & 1 & 12.3025 & 219.1066 & 70.0954 & 15.46 & 270.67 & 0.5568 \\
\hline
FRB 20220914A & 282.0568 & 73.3369 & 0.1139 & 631.05 & 190206 & 11 & 13.0391 & 282.4460 & 73.3689 & 375.76 & 780.52 & 0.0442 \\
 &  &  &  &  & 9333801 & 1 & 11.0119 & 282.0425 & 73.3359 & 22.67 & 129.94 & 0.0761 \\
\hline
FRB 20221012A & 280.7987 & 70.5242 & 0.2847 & 442.20 & 47589 & 21 & 13.7259 & 280.4668 & 70.3779 & 929.47 & 1288.81 & 0.0713 \\
 &  &  &  &  & 1504843 & 4 & 13.4184 & 280.7312 & 70.4544 & 524.30 & 987.09 & 0.1047 \\
\hline
FRB 20221027A & 130.8720 & 72.1009 & 0.5422 & 452.50 & 7048041 & 2 & 11.8690 & 130.9473 & 72.0943 & 279.50 & 279.84 & 0.1864 \\
 &  &  &  &  & 761917 & 5 & 12.5665 & 130.9009 & 72.1316 & 397.67 & 470.99 & 0.2040 \\
 &  &  &  &  & 16926630 & 1 & 11.6420 & 130.8913 & 72.0965 & 119.15 & 187.63 & 0.2877 \\
 &  &  &  &  & 16926665 & 1 & 11.9905 & 130.8634 & 72.1100 & 206.25 & 222.88 & 0.4685 \\
\hline
FRB 20221029A & 141.9634 & 72.4523 & 0.9750 & 1391.05 & 3627804 & 2 & 12.2225 & 141.8244 & 72.3218 & 397.29 & 419.01 & 0.0393 \\
 &  &  &  &  & 3633581 & 2 & 11.8151 & 142.2421 & 72.4806 & 259.16 & 306.38 & 0.0397 \\
 &  &  &  &  & 369720 & 8 & 13.5327 & 141.6510 & 72.5667 & 704.32 & 1115.99 & 0.0667 \\
 &  &  &  &  & 339 & 156 & 14.7755 & 141.9934 & 72.3141 & 1273.04 & 2710.11 & 0.1402 \\
 &  &  &  &  & 529278 & 6 & 13.0882 & 141.9127 & 72.4759 & 413.66 & 674.68 & 0.2544 \\
 &  &  &  &  & 17761869 & 1 & 11.9749 & 141.9662 & 72.4592 & 115.39 & 240.12 & 0.3024 \\
 &  &  &  &  & 17761853 & 1 & 11.6066 & 141.9440 & 72.4537 & 113.46 & 175.26 & 0.3655 \\
 &  &  &  &  & 22422 & 30 & 14.3650 & 142.0382 & 72.3989 & 1186.59 & 1591.01 & 0.4173 \\
 &  &  &  &  & 474422 & 7 & 13.5576 & 142.0350 & 72.4559 & 470.59 & 833.89 & 0.4551 \\
 &  &  &  &  & 842432 & 5 & 13.2722 & 141.8862 & 72.4465 & 527.13 & 661.98 & 0.4724 \\
 &  &  &  &  & 17761850 & 1 & 12.5255 & 141.9396 & 72.4624 & 279.10 & 330.94 & 0.4959 \\
\hline
FRB 20221106A & 56.7048 & -25.5698 & 0.2044 & 343.80 & 5849449 & 2 & 11.4945 & 56.6894 & -25.5707 & 125.61 & 219.03 & 0.1371 \\
\hline
FRB 20221219A & 257.6298 & 71.6268 & 0.5530 & 706.71 & 14626496 & 1 & 11.4937 & 257.5277 & 71.6247 & 159.84 & 188.43 & 0.0697 \\
 &  &  &  &  & 6354 & 52 & 14.3374 & 257.7588 & 71.5778 & 584.42 & 1936.89 & 0.1398 \\
 &  &  &  &  & 489969 & 7 & 12.7595 & 257.6983 & 71.6112 & 316.00 & 552.01 & 0.1913 \\
 &  &  &  &  & 799619 & 5 & 12.9456 & 257.6612 & 71.6088 & 371.90 & 565.30 & 0.3419 \\
 &  &  &  &  & 6764706 & 2 & 12.4189 & 257.6560 & 71.6208 & 218.52 & 349.43 & 0.4490 \\
\hline
FRB 20230124A & 231.9170 & 70.9681 & 0.0939 & 590.00 & 12113008 & 1 & 10.8773 & 231.9092 & 70.9848 & 32.81 & 120.47 & 0.0259 \\
 &  &  &  &  & 12096201 & 1 & 12.2941 & 231.6744 & 70.9499 & 334.11 & 349.50 & 0.0570 \\
 &  &  &  &  & 700875 & 5 & 11.7395 & 232.0271 & 70.9553 & 161.45 & 283.89 & 0.0588 \\
\hline
FRB 20230216A & 156.4722 & 3.4368 & 0.5310 & 828.29 & 927972 & 5 & 13.0782 & 156.4871 & 3.4021 & 471.94 & 697.20 & 0.2046 \\
 &  &  &  &  & 63413 & 18 & 13.8370 & 156.4304 & 3.4128 & 701.80 & 1201.73 & 0.2512 \\
 &  &  &  &  & 39831605 & 1 & 11.7300 & 156.4679 & 3.4337 & 99.46 & 192.71 & 0.3644 \\
\hline
FRB 20230307A & 177.7816 & 71.6951 & 0.2706 & 610.15 & 25543954 & 1 & 11.6390 & 177.7811 & 71.7098 & 116.72 & 205.07 & 0.1184 \\
 &  &  &  &  & 25543953 & 1 & 11.2519 & 177.8163 & 71.6950 & 114.83 & 148.73 & 0.1652 \\
 &  &  &  &  & 498 & 138 & 14.6879 & 177.8819 & 71.5652 & 1580.38 & 2426.42 & 0.1907 \\
 &  &  &  &  & 51814 & 20 & 13.9684 & 177.7470 & 71.7417 & 573.51 & 1392.58 & 0.1943 \\
\hline
FRB 20230616A & 184.6146 & 28.3566 & 0.4485 & 918.33 & 44730822 & 1 & 10.8887 & 184.6898 & 28.3729 & 115.87 & 121.74 & 0.0226 \\
 &  &  &  &  & 44722871 & 1 & 10.9708 & 184.6064 & 28.3891 & 118.00 & 127.81 & 0.0487 \\
 &  &  &  &  & 44722835 & 1 & 11.8799 & 184.6253 & 28.3810 & 163.68 & 250.26 & 0.0903 \\
 &  &  &  &  & 3478192 & 2 & 12.3276 & 184.6353 & 28.3306 & 219.98 & 428.80 & 0.1009 \\
 &  &  &  &  & 838259 & 5 & 12.7266 & 184.5258 & 28.3580 & 563.61 & 579.86 & 0.1058 \\
 &  &  &  &  & 44730716 & 1 & 11.5368 & 184.6058 & 28.3550 & 65.59 & 189.05 & 0.1244 \\
 &  &  &  &  & 3478205 & 2 & 11.9530 & 184.6030 & 28.3558 & 132.87 & 291.35 & 0.2154 \\
 &  &  &  &  & 44730731 & 1 & 12.1300 & 184.6208 & 28.3520 & 110.84 & 274.80 & 0.2716 \\
\hline
FRB 20230626A & 235.6296 & 71.1335 & 0.3270 & 451.20 & 12143965 & 1 & 11.3758 & 235.5629 & 71.1132 & 129.16 & 173.05 & 0.0606 \\
 &  &  &  &  & 3408695 & 2 & 11.7618 & 235.6244 & 71.1219 & 147.01 & 253.82 & 0.2047 \\
 &  &  &  &  & 12120739 & 1 & 11.5972 & 235.5858 & 71.1338 & 187.01 & 188.01 & 0.2200 \\
\hline
FRB 20230628A & 166.7867 & 72.2818 & 0.1270 & 345.15 & 24428027 & 1 & 10.8074 & 167.0642 & 72.2930 & 112.17 & 114.73 & 0.0174 \\
 &  &  &  &  & 5303682 & 2 & 11.4990 & 166.8036 & 72.2414 & 161.13 & 236.98 & 0.0546 \\
\hline
FRB 20230702A & 132.1247 & 26.3895 & 0.3658 & 598.88 & 6527973 & 2 & 11.0240 & 132.0928 & 26.4178 & 54.75 & 170.50 & 0.0180 \\
 &  &  &  &  & 42962969 & 1 & 11.4999 & 132.1199 & 26.4008 & 44.50 & 191.28 & 0.0506 \\
\hline
FRB 20230808F & 53.3041 & -51.9353 & 0.3472 & 653.20 & 648863 & 6 & 12.3818 & 53.2515 & -51.8594 & 367.19 & 463.39 & 0.0620 \\
 &  &  &  &  & 35248386 & 1 & 11.6800 & 53.3046 & -51.9501 & 185.57 & 201.88 & 0.2053 \\
\hline
FRB 20230902A & 52.1398 & -47.3335 & 0.3619 & 440.10 & 35849535 & 1 & 11.4140 & 52.1869 & -47.3115 & 171.53 & 178.07 & 0.0616 \\
 &  &  &  &  & 127915 & 13 & 13.7173 & 52.0889 & -47.4287 & 884.76 & 1211.57 & 0.1321 \\
 &  &  &  &  & 2731302 & 3 & 12.3060 & 52.1360 & -47.3535 & 349.88 & 351.29 & 0.3217 \\
\hline
FRB 20230907D & 187.1425 & 8.6581 & 0.4638 & 1030.79 & 43732346 & 1 & 10.8737 & 187.1608 & 8.6231 & 84.70 & 119.95 & 0.0288 \\
 &  &  &  &  & 477 & 140 & 14.6582 & 186.8694 & 8.8394 & 2086.36 & 2584.44 & 0.0927 \\
 &  &  &  &  & 10701 & 42 & 14.0085 & 187.2688 & 8.7165 & 886.12 & 1569.35 & 0.0929 \\
 &  &  &  &  & 43723269 & 1 & 11.5753 & 187.1358 & 8.6778 & 158.85 & 195.90 & 0.1130 \\
 &  &  &  &  & 3727532 & 2 & 11.7598 & 187.1678 & 8.6409 & 259.25 & 270.52 & 0.1286 \\
 &  &  &  &  & 43723237 & 1 & 12.4505 & 187.1301 & 8.6325 & 357.75 & 363.32 & 0.2071 \\
 &  &  &  &  & 43723267 & 1 & 11.7709 & 187.1360 & 8.6673 & 179.13 & 207.40 & 0.2846 \\
\hline
FRB 20230923A & 206.6141 & 45.4080 & 0.4445 & 918.33 & 6363 & 52 & 14.1245 & 206.7055 & 45.7127 & 1467.45 & 1758.78 & 0.0660 \\
 &  &  &  &  & 4000243 & 2 & 11.5391 & 206.6094 & 45.3852 & 186.96 & 229.89 & 0.1211 \\
 &  &  &  &  & 2540959 & 3 & 12.7197 & 206.6329 & 45.4282 & 333.97 & 516.79 & 0.2342 \\
 &  &  &  &  & 22594453 & 1 & 11.7240 & 206.6057 & 45.3980 & 164.84 & 204.71 & 0.2420 \\
 &  &  &  &  & 22594455 & 1 & 12.0008 & 206.6062 & 45.3991 & 196.00 & 237.62 & 0.3598 \\
\hline
FRB 20231020B & 57.2782 & -37.7699 & 0.4775 & 952.20 & 5803880 & 2 & 12.1975 & 57.2671 & -37.7828 & 198.43 & 353.32 & 0.2091 \\
 &  &  &  &  & 35710502 & 1 & 11.8508 & 57.2767 & -37.7540 & 225.30 & 225.65 & 0.2411 \\
 &  &  &  &  & 35710512 & 1 & 11.7275 & 57.2853 & -37.7735 & 95.33 & 205.01 & 0.2442 \\
 &  &  &  &  & 35710510 & 1 & 11.8502 & 57.2872 & -37.7737 & 134.14 & 218.35 & 0.3019 \\
 &  &  &  &  & 2444794 & 3 & 12.7882 & 57.2442 & -37.7742 & 488.86 & 502.21 & 0.3386 \\
 &  &  &  &  & 5803879 & 2 & 12.2301 & 57.2733 & -37.7651 & 113.63 & 323.65 & 0.3534 \\
 &  &  &  &  & 35710514 & 1 & 12.0518 & 57.2844 & -37.7778 & 179.64 & 244.63 & 0.3788 \\
\hline
FRB 20231102A & 228.8374 & 56.7992 & 0.6878 & 906.20 & 5123186 & 2 & 12.5251 & 228.8000 & 56.7967 & 324.52 & 429.57 & 0.2788 \\
 &  &  &  &  & 8184353 & 1 & 12.5087 & 228.8163 & 56.8071 & 323.06 & 323.15 & 0.5177 \\
\hline
FRB 20231123B & 242.5382 & 70.7851 & 0.2621 & 396.70 & 4048515 & 2 & 11.7383 & 242.4791 & 70.8025 & 224.97 & 265.78 & 0.1299 \\
 &  &  &  &  & 12855832 & 1 & 12.2381 & 242.5403 & 70.7858 & 12.83 & 309.16 & 0.2057 \\
\hline
FRB 20240114A & 321.9160 & 4.3294 & 0.1303 & 527.65 & 31996437 & 1 & 12.5610 & 321.9316 & 4.3899 & 375.85 & 421.76 & 0.0862 \\
 &  &  &  &  & 2467 & 76 & 14.4111 & 321.8629 & 4.4431 & 806.53 & 2137.57 & 0.0929 \\
\hline
FRB 20240215A & 268.4413 & 70.2324 & 0.2100 & 549.50 & 6265153 & 2 & 11.7275 & 268.4862 & 70.2775 & 200.42 & 281.40 & 0.0584 \\
\hline
FRB 20240304B & 182.9970 & 11.8130 & 2.1480 & 2458.20 & 45957221 & 1 & 11.5284 & 182.9961 & 11.8251 & 161.20 & 178.17 & 0.2221 \\
 &  &  &  &  & 126478 & 13 & 13.8394 & 183.0109 & 11.8196 & 252.66 & 1153.87 & 0.3055 \\
 &  &  &  &  & 1369603 & 4 & 13.0039 & 182.9864 & 11.8254 & 351.98 & 541.79 & 0.4642 \\
 &  &  &  &  & 3093571 & 2 & 13.4217 & 182.9906 & 11.8289 & 480.94 & 597.44 & 0.8298 \\
\hline
\cutinhead{Isolated-galaxy-only sight lines ($R=1$ only)}
FRB 20211212A & 157.3508 & 1.3604 & 0.0707 & 209.00 & 40043548 & 1 & 11.3907 & 157.3035 & 1.3469 & 153.15 & 176.75 & 0.0424 \\
\hline
FRB 20220310F & 134.7205 & 73.4908 & 0.4780 & 462.15 & 16925867 & 1 & 10.8769 & 134.8890 & 73.5016 & 86.28 & 120.60 & 0.0234 \\
\hline
FRB 20220330D & 163.7512 & 70.3508 & 0.3714 & 468.10 & 24478365 & 1 & 11.1418 & 163.7387 & 70.3616 & 62.59 & 143.50 & 0.0762 \\
\hline
FRB 20220509G & 282.6700 & 70.2438 & 0.0894 & 269.50 & 9332678 & 1 & 10.8587 & 282.6266 & 70.2563 & 53.60 & 118.02 & 0.0377 \\
\hline
FRB 20230222B & 238.7391 & 30.8987 & 0.1100 & 188.03 & 37036773 & 1 & 11.6692 & 238.7753 & 30.8874 & 176.56 & 214.76 & 0.0757 \\
 &  &  &  &  & 37036798 & 1 & 11.6264 & 238.7165 & 30.9252 & 176.43 & 207.80 & 0.0761 \\
\hline
FRB 20230526A & 22.2326 & -52.7173 & 0.1570 & 361.40 & 10972428 & 1 & 10.9927 & 22.2009 & -52.7109 & 54.56 & 130.83 & 0.0364 \\
 &  &  &  &  & 10972405 & 1 & 11.1382 & 22.2528 & -52.7264 & 119.86 & 139.97 & 0.1170 \\
\hline
FRB 20230712A & 167.3585 & 72.5578 & 0.4525 & 586.96 & 24429101 & 1 & 11.6039 & 167.3941 & 72.5561 & 132.70 & 190.97 & 0.2005 \\
\hline
FRB 20231220A & 123.9087 & 73.6599 & 0.3355 & 491.20 & 19934090 & 1 & 11.5308 & 123.8632 & 73.6717 & 156.32 & 186.96 & 0.1365 \\
 &  &  &  &  & 19934094 & 1 & 11.3639 & 123.8945 & 73.6617 & 44.78 & 162.53 & 0.1595 \\
 &  &  &  &  & 19934103 & 1 & 11.7299 & 123.9103 & 73.6597 & 7.46 & 203.56 & 0.2609 \\
\hline
FRB 20240210A & 8.7796 & -28.2708 & 0.0237 & 283.73 & 18502019 & 1 & 10.9422 & 8.7840 & -28.3009 & 51.01 & 126.84 & 0.0223 \\
\hline
FRB 20240213A & 166.1683 & 74.0754 & 0.1185 & 357.40 & 24403782 & 1 & 10.8808 & 166.0673 & 74.0612 & 99.80 & 119.64 & 0.0437 \\
\hline
FRB 20240310A & 17.6219 & -44.4394 & 0.1270 & 601.80 & 24484233 & 1 & 11.4630 & 17.7039 & -44.4895 & 166.20 & 188.15 & 0.0289 \\
\hline
FRB 20241228A & 216.3858 & 12.0250 & 0.1614 & 246.53 & 28958311 & 1 & 11.8393 & 216.3850 & 12.0225 & 17.59 & 241.81 & 0.0965 \\
\hline
\cutinhead{Clean nonintersecting sight lines}
FRB 20180924B & 326.1052 & -40.9000 & 0.3212 & 362.16 &  &  &  &  &  &  &  &  \\
FRB 20181119A & 190.4799 & 65.1119 & 0.2606 & 366.10 &  &  &  &  &  &  &  &  \\
FRB 20190425A & 255.6750 & 21.5764 & 0.0312 & 128.20 &  &  &  &  &  &  &  &  \\
FRB 20190608B & 334.0199 & -7.8982 & 0.1178 & 340.05 &  &  &  &  &  &  &  &  \\
FRB 20200120E & 149.4778 & 68.8189 & 0.0008 & 87.82 &  &  &  &  &  &  &  &  \\
FRB 20210603A & 10.2741 & 21.2263 & 0.1772 & 500.15 &  &  &  &  &  &  &  &  \\
FRB 20220529A & 19.1041 & 20.6324 & 0.1839 & 246.40 &  &  &  &  &  &  &  &  \\
FRB 20220920A & 240.2571 & 70.9188 & 0.1582 & 315.00 &  &  &  &  &  &  &  &  \\
FRB 20230203A & 151.6616 & 35.6941 & 0.1464 & 417.30 &  &  &  &  &  &  &  &  \\
FRB 20230708A & 303.1155 & -55.3563 & 0.1050 & 411.51 &  &  &  &  &  &  &  &  \\
FRB 20230918A & 161.0507 & 61.1209 & 0.0600 & 365.96 &  &  &  &  &  &  &  &  \\
FRB 20230926A & 269.1249 & 41.8143 & 0.0553 & 222.81 &  &  &  &  &  &  &  &  \\
FRB 20231005A & 246.0280 & 35.4487 & 0.0713 & 188.84 &  &  &  &  &  &  &  &  \\
FRB 20231120A & 143.9840 & 73.2847 & 0.0368 & 438.90 &  &  &  &  &  &  &  &  \\
FRB 20231223C & 259.5446 & 29.4979 & 0.1059 & 179.54 &  &  &  &  &  &  &  &  \\
FRB 20231226A & 155.3637 & 6.1103 & 0.1569 & 329.90 &  &  &  &  &  &  &  &  \\
FRB 20240201A & 149.9056 & 14.0880 & 0.0427 & 374.50 &  &  &  &  &  &  &  &  \\
FRB 20250316A & 182.4347 & 58.8491 & 0.0065 & 161.82 &  &  &  &  &  &  &  &  \\
\enddata
\tablenotetext{1}{Group identifier in the Y21/DESI DR9 group catalog.}
\tablenotetext{2}{$\log_{10}(M_h/[h^{-1}M_\odot])$ from the Y21 group catalog.}
\tablenotetext{3}{Projected impact parameter between the FRB sight line and the foreground-system center.}
\tablenotetext{4}{For systems with richness $R>1$, $R_{\rm halo}=r_{180}$ using the Y21 definition adopted in this work. For isolated systems with richness $R=1$, $R_{\rm halo}$ denotes the converted virial radius of the galaxy halo.}
\tablecomments{This table lists the 61 FRB sight lines inside the Y21/DESI group-catalog footprint: 31 group-intersecting sight lines, 12 isolated-galaxy-only sight lines, and 18 clean nonintersecting sight lines.}
\end{deluxetable}
\end{longrotatetable}

\clearpage
\onecolumngrid

\setcounter{figure}{0}
\renewcommand{\thefigure}{C\arabic{figure}}
\renewcommand{\theHfigure}{C\arabic{figure}}
\setcounter{table}{0}
\renewcommand{\thetable}{C\arabic{table}}
\renewcommand{\theHtable}{C\arabic{table}}
\setcounter{equation}{0}
\renewcommand{\theequation}{C\arabic{equation}}
\renewcommand{\theHequation}{C\arabic{equation}}

\section{{Statistical Model Diagnostics and Sensitivity Tests}}
\label{app:statistical-diagnostics}

\subsection{{Joint Posterior Structure}}
\label{app:joint-posterior}

\begin{figure}[ht]
    \centering
    \includegraphics[width=0.86\linewidth]{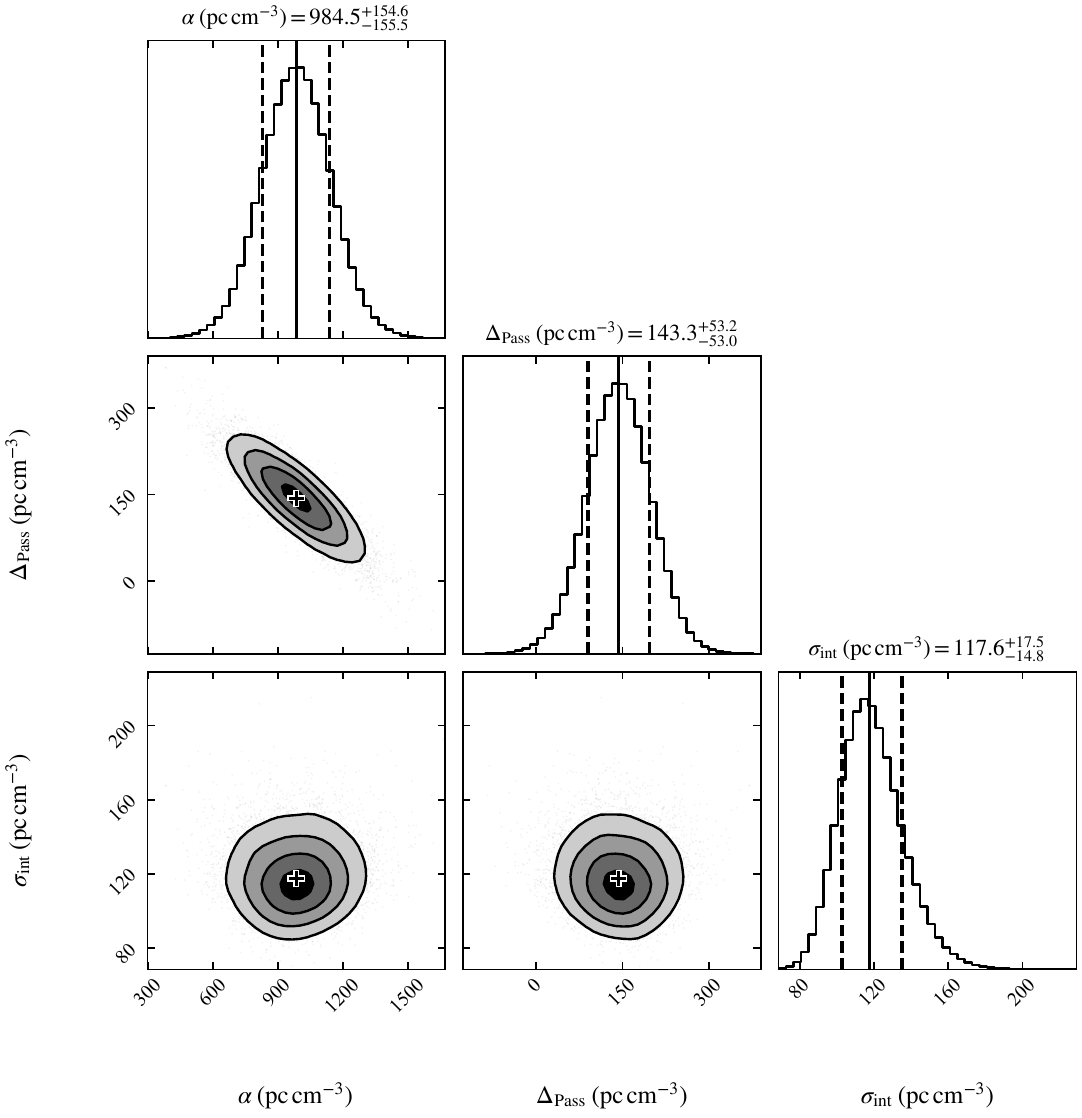}
    \caption{{Joint posterior distributions of $\alpha$, $\Delta_{\rm Pass}$, and $\sigma_{\rm int}$ obtained from the joint Pass--NoPass fit. Diagonal panels show the marginal posterior distributions, with dashed lines marking the 16th and 84th percentiles and solid lines marking the medians. Off-diagonal panels show contours with enclosed probabilities equivalent to $0.5$, $1.0$, $1.5$, and $2.0\,\sigma$ for a bivariate Gaussian; black plus signs mark the corresponding pairs of marginal posterior medians.}}
    \label{fig:joint-offset-corner}
\end{figure}

{Figure~\ref{fig:joint-offset-corner} shows the joint posterior distributions of $\alpha$, $\Delta_{\rm Pass}$, and $\sigma_{\rm int}$ obtained from the Pass--NoPass fit presented in Section~\ref{sec:statistical-analysis}.}

{The posterior is unimodal and shows no evidence of truncation at the adopted prior boundaries. The strongest parameter dependence is the anticorrelation between $\alpha$ and $\Delta_{\rm Pass}$ ($\rho=-0.860$): a steeper common redshift relation requires a smaller Pass-sample offset. This covariance is included in the marginalized uncertainty reported for $\Delta_{\rm Pass}$.}

\subsection{{Sensitivity to the Residual-Scatter Model}}
\label{app:separate-scatter}

{In the primary joint-offset model, the common $\sigma_{\rm int}$ describes residual variation about the fitted mean relation beyond the individual $\sigma_{{\rm cosmic},i}$ uncertainties. We test whether the inferred mean offset changes when the Pass and NoPass samples are modeled with separate residual scatters, retaining the same mean relation, $\mu_i=\alpha z_i+\Delta_{\rm Pass}P_i$. The variance is then}
\begin{equation}
s_i^2
=
\sigma_{{\rm cosmic},i}^2
+
(1-P_i)\sigma_{\rm int,NoPass}^2
+
P_i\sigma_{\rm int,Pass}^2 .
\label{eq:separate-scatter}
\end{equation}

{We retain the priors on $\alpha$ and $\Delta_{\rm Pass}$ used in the primary analysis and assign the same log-uniform prior independently to $\sigma_{\rm int,NoPass}$ and $\sigma_{\rm int,Pass}$. We use the same sampling configuration as in the primary analysis, except that the ensemble is run for 12,000 steps. The resulting postburn chain length exceeds 50 times the maximum integrated autocorrelation time among the model parameters.}

{The separate-scatter fit gives}
\begin{equation}
\begin{aligned}
\alpha
&=
896.9^{+124.3}_{-101.5}
~\mathrm{pc\,cm^{-3}},\\
\Delta_{\rm Pass}
&=
167.0^{+46.3}_{-48.4}
~\mathrm{pc\,cm^{-3}},\\
\sigma_{\rm int,NoPass}
&=
38.1^{+32.6}_{-38.0}
~\mathrm{pc\,cm^{-3}},\\
\sigma_{\rm int,Pass}
&=
145.2^{+28.9}_{-22.7}
~\mathrm{pc\,cm^{-3}}.
\end{aligned}
\end{equation}

{The posterior gives $P(\Delta_{\rm Pass}>0)=0.9991$. The inferred $\Delta_{\rm Pass}$ is consistent with the primary common-scatter result, indicating that the positive mean offset persists when separate residual scatters are allowed.}

\subsection{{Complementary Statistical Comparisons of the Pass and NoPass Samples}}
\label{app:complementary-tests}

{We apply complementary regression and distributional comparisons to the same 23 Pass and 18 NoPass sight lines. We set $y_i$ to the posterior median of $\mathrm{DM}_{{\rm cosmic},i}$ and use the symmetrized uncertainty $\sigma_{{\rm cosmic},i}$ defined in Equation~\ref{eq:sigma-cosmic}.}

\subsubsection{{Comparison of the Regression Coefficients}}

{We apply a Chow test modified to incorporate the individual uncertainties in the inferred $\mathrm{DM}_{{\rm cosmic},i}$ values \citep{Chow1960Econometrica..28..591C}. We compare a restricted model in which the Pass and NoPass samples share a common intercept and slope with an unrestricted model that allows both coefficients to differ between the samples:}
\begin{equation}
\begin{aligned}
y_i &= b_0+b_1z_i+\epsilon_i,\\
y_i &= b_0+b_1z_i+c_0P_i+c_1z_iP_i+\epsilon_i,
\end{aligned}
\end{equation}
{respectively. Here, $c_0$ and $c_1$ represent the Pass--NoPass differences in intercept and slope, respectively.}

{We define the total variance of each sight line as}
\begin{equation}
v_i=
\begin{cases}
\sigma_{{\rm cosmic},i}^2+\widehat{\sigma}_{\rm int,0}^2,
& \text{shared scatter},\\
\sigma_{{\rm cosmic},i}^2
+(1-P_i)\widehat{\sigma}_{\rm int,NoPass,0}^2
+P_i\widehat{\sigma}_{\rm int,Pass,0}^2,
& \text{separate scatters}.
\end{cases}
\end{equation}
{The subscript 0 denotes values estimated by maximum likelihood under the corresponding restricted model. For each variance specification, the same weights $v_i^{-1}$ are used in the restricted and unrestricted fits. Their weighted residual sums of squares are}
\begin{equation}
Q_m
=
\sum_i
\frac{(y_i-\mu_{m,i})^2}{v_i},
\qquad
m\in\{{\rm R},{\rm U}\},
\end{equation}
{where $\mu_{m,i}$ is the fitted mean, and ${\rm R}$ and ${\rm U}$ denote the restricted and unrestricted models defined above, respectively. For $N=41$, the test statistic is}
\begin{equation}
F_W
=
\frac{(Q_{\rm R}-Q_{\rm U})/2}
{Q_{\rm U}/(N-4)} .
\end{equation}

{Because the weights depend on fitted scatter parameters, we calibrate $F_W$ separately for the two variance specifications using 100,000 parametric-bootstrap realizations. In each realization, mock values are drawn as $y_i^{*}\sim\mathcal{N}(\widehat{\mu}_{{\rm R},i},v_i)$, while the redshifts, sample labels, and individual $\mathrm{DM}_{\rm cosmic}$ uncertainties are held fixed. The restricted model is then refitted to update its mean, scatter parameters, and weights, and both models are refitted to obtain $F_W^{*}$. We calculate $p_{\rm boot}=(1+n_{\geq})/(100{,}000+1)$, where $n_{\geq}$ is the number of bootstrap statistics at least as large as the observed value.}

\subsubsection{{Intercept Comparison under a Common Slope}}

{Following \citet{FeigelsonBabu1992ApJ...397...55F}, we estimate a slope $\widehat{b}_{\rm C}$ by unweighted least squares from one calibration sample and apply it to both samples. The Pass--NoPass intercept difference, expressed in $\mathrm{pc\,cm^{-3}}$, is}
\begin{equation}
\widehat{\Delta}
=
\left(
\overline{y}_{\rm Pass}
-
\overline{y}_{\rm NoPass}
\right)
-
\widehat{b}_{\rm C}
\left(
\overline{z}_{\rm Pass}
-
\overline{z}_{\rm NoPass}
\right).
\end{equation}
{We calculate its uncertainty using the influence-function method of \citet{FeigelsonBabu1992ApJ...397...55F} and obtain two-sided $p$-values from the normal approximation. Because the result depends on the calibration sample, we evaluate both calibration directions.}

\subsubsection{{Comparison of the Slope-Removed Distributions}}

{To examine whether the Pass and NoPass samples differ in their overall distributions, including their widths and tails, we apply Anderson--Darling tests after removing the fitted common redshift trend:}
\begin{equation}
R_i
=
y_i-\widehat{\alpha}z_i ,
\end{equation}
{where $\widehat{\alpha}=982.764~\mathrm{pc\,cm^{-3}}$ is the maximum-likelihood slope from the shared-scatter offset fit. We first calculate the unstandardized Scholz--Stephens two-sample midrank Anderson--Darling statistic with equal weights \citep{ScholzStephens1987JASA...82..918S}.}

{As a precision-weighted sensitivity test, we also apply the weighted Anderson--Darling extension of \citet{Linden2026arXiv260721782L}, assigning each sight line the weight}
\begin{equation}
w_i
=
\left(
\sigma_{{\rm cosmic},i}^2
+
\widehat{\sigma}_{\rm int}^2
\right)^{-1},
\end{equation}
{where $\widehat{\sigma}_{\rm int}=113.456~\mathrm{pc\,cm^{-3}}$. For group $g\in\{0,1\}$, corresponding to NoPass and Pass, respectively, the weighted empirical cumulative distribution and Kish effective sample size are}
\begin{equation}
\begin{aligned}
\widehat{F}_g^{\,w}(r)
&=
\frac{
\sum_{i:P_i=g}w_i\,\mathbb{I}(R_i\leq r)
}{
\sum_{i:P_i=g}w_i
},\\
n_{e,g}
&=
\frac{
\left(\sum_{i:P_i=g}w_i\right)^2
}{
\sum_{i:P_i=g}w_i^2
}.
\end{aligned}
\end{equation}
{The pooled weighted distribution and total effective sample size are}
\begin{equation}
\widehat{D}^{\,w}(r)
=
\frac{
\sum_i w_i\,\mathbb{I}(R_i\leq r)
}{
\sum_i w_i
},
\qquad
n_e=n_{e,0}+n_{e,1}.
\end{equation}
{For the ordered distinct residuals $R_{(1)}<\cdots<R_{(M)}$, the weighted statistic is}
\begin{equation}
A_w^2
=
\sum_{k=1}^{M-1}
\frac{
\left[
\widehat{F}_1^{\,w}(R_{(k)})
-
\widehat{F}_0^{\,w}(R_{(k)})
\right]^2
}{
2\widehat{D}^{\,w}(R_{(k)})
\left[1-\widehat{D}^{\,w}(R_{(k)})\right]/n_e
}
\,\tau_k ,
\end{equation}
{where $\tau_k$ is the number of sight lines sharing $R_{(k)}$. Both Anderson--Darling $p$-values are calibrated using 100,000 label permutations that preserve the Pass and NoPass sample sizes. For each permutation, the sample labels are reassigned, the shared-scatter offset model is refitted, and $R_i$, $w_i$, and both statistics are recomputed. We calculate $p=(1+n_{\geq})/(100{,}000+1)$, where $n_{\geq}$ is the number of permuted statistics at least as large as the observed value. Table~\ref{tab:statistical-tests} summarizes the results.}

\begin{deluxetable}{lcc}
\tabletypesize{\scriptsize}
\tablecaption{{Complementary statistical comparisons of the Pass and NoPass samples.}\label{tab:statistical-tests}}
\tablehead{
\colhead{{Method}} &
\colhead{{Statistic or offset}} &
\colhead{{$p$-value}}
}
\startdata
{Weighted Chow: shared scatter} &
{$F_W=3.810$} &
{0.0315} \\
{Weighted Chow: separate scatters} &
{$F_W=5.752$} &
{0.00576} \\
{Feigelson--Babu: NoPass calibration} &
{$\widehat{\Delta}=195.6\pm49.2$} &
{$7.09\times10^{-5}$} \\
{Feigelson--Babu: Pass calibration} &
{$\widehat{\Delta}=104.1\pm63.4$} &
{0.101} \\
{Scholz--Stephens midrank A--D} &
{$A_{\rm raw}^2=3.017$} &
{0.0272} \\
{Weighted A--D: total-variance weights} &
{$A_w^2=283.084$} &
{0.0158} \\
\enddata
\end{deluxetable}
\onecolumngrid

{Both weighted Chow specifications and both Anderson--Darling comparisons disfavor their respective null hypotheses at the 5\% level. The Feigelson--Babu significance depends on the calibration direction. The statistical support for a Pass--NoPass difference is therefore method-dependent.}
\end{document}